\documentclass[4paper,fleqn]{cas-sc}

\usepackage[authoryear,longnamesfirst]{natbib}

\usepackage{xcolor}
\usepackage[english]{babel}
\usepackage{graphicx}
\usepackage{enumerate}
\usepackage{tikz}
\usepackage[left]{lineno}

\usepackage{amsmath}
\usepackage{float}
\usepackage{lscape}

\makeatletter
\DeclareRobustCommand{\V}{\text{\volumedash}V}
\newcommand{\volumedash}{%
  \makebox[0pt][l]{%
    \ooalign{\hfil\hphantom{$\m@th V$}\hfil\cr\kern0.08em--\hfil\cr}%
  }%
}
\makeatother
\renewcommand{\d}{\textrm{d}}

\newcommand{\sep}{,}

\begin{document}
\let\WriteBookmarks\relax
\def\floatpagepagefraction{1}
\def\textpagefraction{.001}
\shorttitle{Transient stratification force on particles crossing a density interface}
\shortauthors{Verso \emph{et al.}}

\title [mode = title]{Transient stratification force on particles crossing a density interface}                      

\author[1]{Lilly Verso}
\address[1]{School of Mechanical Engineering, Tel Aviv University, Tel Aviv 69978, Israel}

\author[2]{Maarten van  Reeuwijk}
\address[2]{Department of Civil and Environmental Engineering, Imperial College London, London SW7 2AZ, UK}

\author[1]{Alexander Liberzon}

\cortext[cor1]{corr. author email: lillyverso@gmail.com}
\maketitle

\begin{abstract}
We perform a series of experiments to measure Lagrangian trajectories of settling and rising particles as they traverse a density interface of thickness $h$ using an index-matched water-salt-ethanol solution.
The experiments confirm the substantial deceleration that particles experience as a result of the additional force exerted on the particle due to the sudden change in density.
This stratification force is calculated from the measurement data for all particle trajectories. In absence of suitable parameterisations in the literature, a simple phenomenological model is developed which relies on parameterisations of the effective wake volume and recovery time scale.
The model accurately predicts the particle trajectories obtained in our experiments and those of \cite{Fernando1999}.
Furthermore, the model demonstrates that the problem depends on four key parameters, namely the entrance Reynolds number $Re_1$, entrance Froude number $Fr$, particle to fluid density ratio $\rho_p/\rho_f$, and relative interface thickness $h/a$.


\end{abstract}



\section{Introduction}\label{sec:introduction}

Accurate prediction of settling rates of particles in stratified environments is important for the dispersion of pollutants in the atmosphere \citep{turco1983,kok2011},
accumulation of marine snow \citep{macintyre1995,prairie2013}, and oxygen regulation for ocean organisms by bubbles rising across the ocean thermocline \citep{smith1992,burd2009}.
Particles were observed to reside much longer in stratified layers than in a uniform density fluid \citep{macintyre1995,Camassa2013}. This is caused by the sudden change in fluid density as perceived by the moving particle, and in some cases also surface tension effects, which create an additional drag force, in addition to the gravitational, drag, added mass and Basset forces, present in homogeneous density layers~\citep{geller1986,eames1997,Fernando1999, Magnaudet2020}.

One of the central features of particles crossing density interfaces, layers between two fluids of different density, is that they distort the isopycnals by dragging along fluid from the top layer into the new environment. 
The density difference between the fluid in the particle wake (referred to as the caudal wake) and the ambient fluid results in an additional force on the particle as it crosses the interface~\citep{Fernando1999}. 
The caudal wake and its break-up has been extensively investigated for both immiscible and miscible fluids \citep[][and references therein]{Fernando1999, magnaudet2018, magnaudet2018b}. 
Immiscible interfaces are sharp by definition and are subject to surface tension. Here, ``sharpness'' of the interface is characterised by the ratio of the interface thickness to the particle diameter, $h/a \ll 1$. The focus of this work is on particle settling in an environmental context of finite thickness density interfaces, which implies that the fluids are miscible and the interface is of a finite thickness, $h/a = O(1)$. 
The net effect of the caudal wake is that after the particle has passed, fluid particles have been displaced from their original position, and the integral of the displaced fluid is usually referred to as the  drift volume \citep{Magnaudet2020}.

The first study to characterise in detail the drag induced by a finite thickness density jump is by~\citet[][hereafter abbreviated as SMF]{Fernando1999}. They used a water-alcohol-brine system and considered  particles in the range $1.5<Re<15, 3<Fr<10$, where $Re$ is the entrance Reynolds number and $Fr$ is the entrance Froude number, defined as, respectively, $Re = V a/\nu$ and $Fr = V/Na$.
Here, $V$ is the particle settling velocity, $\nu$ is the kinematic viscosity and $N$ is the Brunt-V\"ais\"al\"a frequency (see \S  \ref{sec:experimental_setup} for definition).
The study convincingly showed that particles pull a caudal column of the fluid from a top (lighter) layer into the interface layer, and distort isopycnals that return to the original positions after particles pass~\citep[see also][]{torres2000,okino2017}.
The drag on the particle was observed to increase tenfold, causing a significant slowdown of particles entering the stratified layer. The particle velocity continued to decrease, until reaching a minimum after which the particle accelerated again. Visualisations indicated that minimum was associated with a rupturing wake.
SMF estimated the stratification force $F_S$ using an integral of the photographed caudal wake volume. 
The volume was assumed to be axisymmetric and its radius was estimated from the photographs, from the moment of entrance till the presumed rupture point. The additional buoyancy of the caudal wake volume was modelled as a drag force, and presented as a drag coefficient extension for the stratified layer case, $C_{DS}$. The authors focused primarily on the drag enhancement and did not investigate in detail the particle motion after
the crossing, in which the particle adjusts to a new steady state velocity. This will be denoted the ``recovery phase''. 
The study of this phase along with the prediction of the total settling time will form a central part of the current work.

\cite{Adalsteinsson2004} performed experiments similar to those of SMF for $h/a=O(1)$ and particles with $ 20 < Re < 400$ and $ 5 < Fr < 20$.
In some cases they observed a temporary reversal of the particle velocity as it entered the density interface, and coined it ``particle levitation''. The authors developed a model of the caudal wake, which indicated that the levitation phenomenon depended critically on the mixing of fluid into the wake.
\cite{Camassa2009, Camassa2010} explored the behaviour of low-Reynolds particles crossing a two-layer stratification ($h/a \ll 1$) created by carefully pouring corn syrup on a layer of corn syrup mixed with salt.
An elegant theoretical model was developed as part of this work, which took advantage of the linearity of the Stokes equation and composing the velocity field into a homogeneous Stokes flow and a perturbation velocity due to the density interface. The latter was determined using a free-space Green's function over the fluid domain. 

Stratification-induced drag enhancement also features in  linearly stratified environments ($h/a \gg 1$). \cite{yick2009} studied the behaviour of low-Reynolds number particles and observed that the caudal wake fluid was continuously replaced, causing a quasi-steady drag. 
The drag was shown to scale as the square root of a viscous Richardson number $Ri_v=a^3 N^2 / (\nu V)$.
With increasing particle Reynolds number, the wake reduces to a single filament that alters the pressure on the rear region of the sphere. This results once again in drag increase, with consequent decrease of settling rate in the stratified layer~\citep[see also][]{zvirin1975,torres2000,ardekani2010}. 
Very recently, \cite{Zhang2019} introduced a rigourous  decomposition technique which splits the drag force into different contributions.
Their results, obtained by performing a set of direct numerical simulations, indicate that the drag enhancement is not generally due to the extra buoyancy force resulting from dragging of light fluid by the body, but rather to the specific structure of the vorticity field created by the buoyancy effects.

The transient dynamics of a sphere, released at rest in a linearly stratified fluid and accelerating to its settling speed, was investigated by \cite{Doostmohammadi2014}. 
A maximum velocity during the early stages of the motion was documented. The authors proposed a numerical model based on a transient stratification drag to predict the peak velocity. 
\cite{Candelier2014} attributed the sudden deceleration experienced by the sphere in the initial stages of the motion to an exceeding memory force. 
It was found that eventually this perturbation force tends to a constant value that is incorporated as a correction term of the steady drag force.

In this paper we study the stratification-induced force on finite Reynolds number particles as they enter, traverse and leave a density interface of finite thickness ($h/a \sim 10)$. We use Particle Tracking Velocimetry (PTV) with the refractive index-matching to obtain accurate information on particle position, velocity and acceleration, particularly in the density interface where deceleration are largest. 
Four different particle types are used, in order to explore in detail the effect of particle density, and size, on the stratification-induced forces. 
The experiments demonstrate that both the particle to fluid ratio, $\rho_p/\rho_1$, and the interface sharpness ratio, $h/a$, are important factors for the stratification force. 
Specifically, we estimate the stratification force from the experimental data, and 
develop a simple model.
We demonstrate that the previously observed minimal velocity and the time to reach minimal velocity coincide with the particle exiting the density interface.  Furthermore, we investigate the time it takes the particles to attain their new terminal velocity (the recovery time) and we compare it with the particle time scale. We conclude the analysis by exploring how the stratification force and the associated crossing and recovery time scales depend on the four dimensionless quantities that govern this problem, namely $Re$, $Fr$, $\rho_p/\rho_f$, and $h/a$.

\section{Experimental details}\label{sec:experimental_setup}

We study the motion of a spherical particle of diameter $a$ and density $\rho_p$ that
crosses a stratified density interface layer of finite thickness ($h$), between two homogeneous layers of fluids of densities $\rho_2 > \rho_1$. In the general case the dynamic viscosity, $\nu_1$ and $\nu_2$ of the two fluid layers are also different.
\begin{figure}
\centering
    \includegraphics[height=6cm]{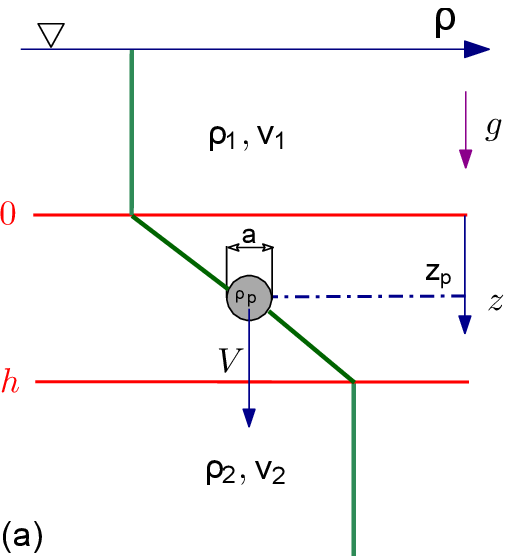}
    ~
\includegraphics[height=6cm,width=3.25cm]{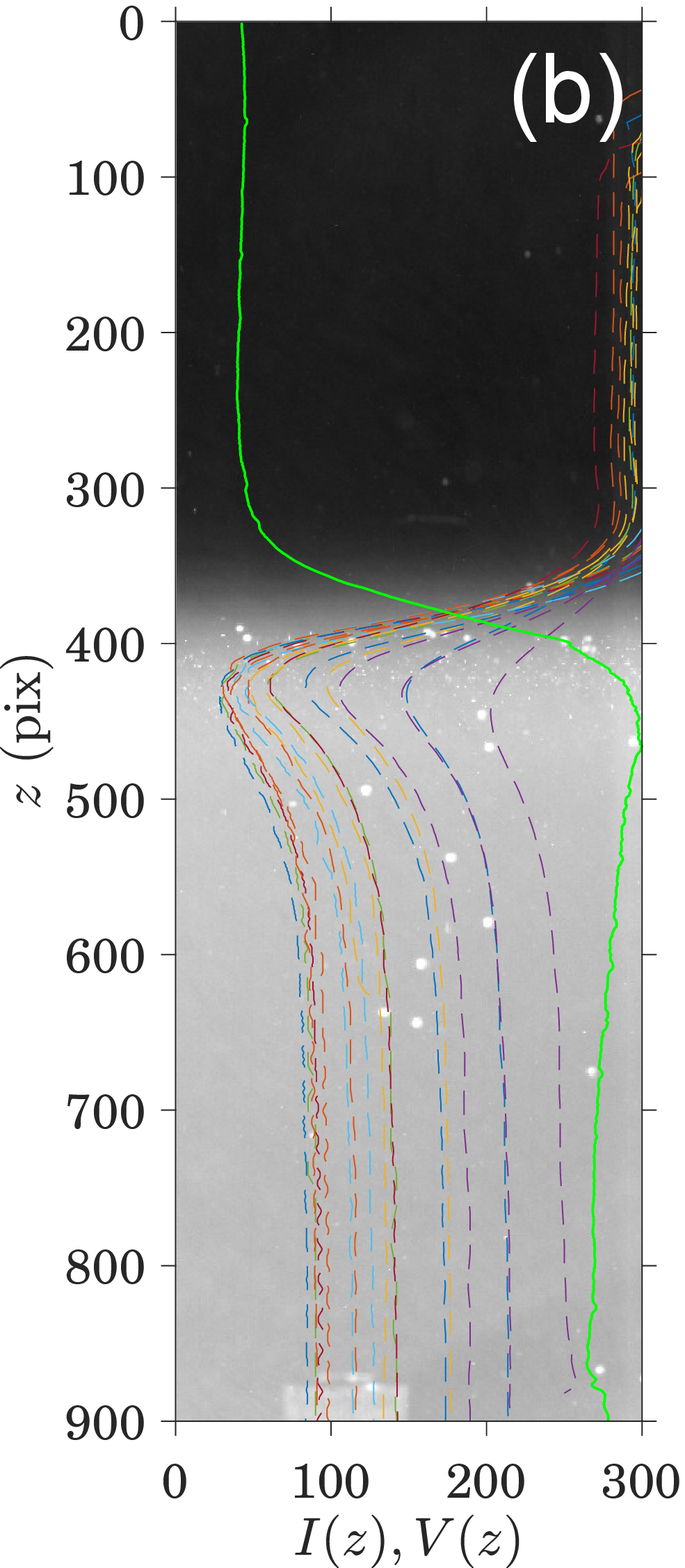}   
\caption{(a) Schematic definition of the problem. (b) Visualisation of the stratification in the two layers using background illumination. Normalised average intensity is presented in a green solid line overlaid with the colour dashed lines of the normalised vertical velocity of particles P1 (see Fig.~\ref{fig:raw_p1p2p3p4} for quantitative measurements).}\label{fig:sketch_problem}
\end{figure}

The experiments were carried out in a glass tank with a 200 $\times$ 200 mm$^{2}$ cross-section and a depth of 300 mm, as that used in the study of \cite{verso2017}. Two series of experiments were performed using a combination of water solutions of ethanol and Epsom salts (MgSO$_4$), with two different concentrations, a lighter solution of ethanol and water and a heavier saline solution of Epsom salts and water. The values of fluid density ($\rho_1,\rho_2$), kinematic viscosity ($\nu_1,\nu_2$) and Brunt-V\"ais\"al\"a frequency ($N$) are shown in table~\ref{tab:properties_layers}, where $N$ is calculated as
\begin{equation}\label{eq:N}
N =\left(\frac{2g}{\rho_1+\rho_2}\frac{\rho_2-\rho_1}{h}\right)^{1/2}
\end{equation}
The working fluid was prepared to ensure that the refractive indexes across the stratified layer were matched, thereby minimising the distortion of the scattered light to the camera \citep[e.g.][among others]{alahyari1994}. Refractive index matching minimises the optical errors due to high density gradients and allows to track particles moving through both layers and the interface.
In order to obtain the two-layer stratified medium as sketched in figure~\ref{fig:sketch_problem}, the light fluid was first introduced into the tank, after which the heavy fluid was pumped slowly from the bottom opening. 
The filling procedure was controlled by a peristaltic pump and the flow rate that was manually adjusted within 10-40 mL/min to minimise mixing of the two layers. 

In experiment 1 the viscosity values of the both layers were measured using a Cannon-Fenske Routine viscometer with an uncertainty of $\pm 17\%$. In experiment 2, the viscosity was not measured, but due to the careful preparation procedure of the fluids in a controlled environment, the values can safely be assumed identical to series 1 within this uncertainty range. All the fluid properties for the two series, along with the measurements of the thickness $h$ and a respective value of $N$, are reported in table~\ref{tab:properties_layers}. 

\begin{table}
\begin{center}
\def~{\hphantom{0}}
\caption{Properties of the fluid layers in experiments 1 and 2.}
\label{tab:properties_layers}
\begin{tabular}{ccccccc}
\hline
& $\rho_1$ & $\rho_2$  & $\nu_1$ & $\nu_2$ & $h$ & $N$ \\
\textbf{Exp}  & (kg m$^{-3})$ & (kg m$^{-3}$) & (m$^{2}$ s$^{-1}$)  & (m$^{2}$ s$^{-1}$) & (m) & (s$^{-1}$) \\ \hline
\textbf{1}     &    976    &   1025     &     $1.43\times 10^{-6}$     &   $1.012\times 10^{-6}$          &  0.013    &  6.08\\ \hline
\textbf{2}     &    975    &     1020   &      $1.43\times 10^{-6}$     &    $1.012\times 10^{-6}$        &   0.01    &  6.4\\ \hline
\end{tabular}
\end{center}
\end{table}

We used four distinctly different types of spherical particles, named P1-P4, as reported in table~\ref{tab:particle_types}. The particles are commercially available from Cospheric (Santa Barbara, CA) with identification numbers WPMS 850-1000 $\mu$m (P1), WPMS 425-500 $\mu$m (P2) and SLGMS 710-850 $\mu$m (P3) and CPMS-0.96 850-1000 $\mu$m (P4). The particles P1, P2, P4 are manufactured in polystyrene and P3 in soda-lime glass, spanning a range of diameters and density ratios, $\rho_p/\rho_1$. Hereafter all the properties related to the top layer will be marked with a subscript $i=1$ and all the properties estimated in the bottom layer will be marked with a subscript $i=2$. 
The tests are performed in a range of $2<Re<106$ and $0.5<Fr<28$ as shown in table 1. Within this range, the particles fall without instability during their descent (\cite{magnaudet2018}).

During experiment 1 (see table~\ref{tab:properties_layers}), individual particles of types P1-P3 were released at the centre of the tank and below of the free surface level, to free fall through the stratified interface. During the  experiment 2, fluids were changed as shown in table~\ref{tab:properties_layers} and individual particles of type P4 were released from the bottom opening of the aquarium into the heavy fluid layer to raise through the stratified interface towards the free surface.
Overall we released and tracked individually 18, 11, 11, and 15 particles of types P1-4, respectively. 
In both series the individual particles were released at sufficiently large time intervals to ensure that the fluid was quiescent again. 

\begin{table}
\def~{\hphantom{0}}
\begin{center}
\caption{Properties of the particles as provided by the manufacturers and dimensionless numbers of top and bottom layer.}
\label{tab:particle_types}
\begin{tabular}{ccccccccc}
\hline
Type & $\rho_p$ (kg m$^{-3}$) & $a$ ($\mu$m)  &  $Re_1$ & $Re_2$ & $Fr_1$ & $Fr_2$ & $\rho_p/\rho_1$ & $h/a$ \\ \hline
\textbf{P1}            & 1033--1100                 & 850--1000      &  6--14    &  2--15      & 2--4   &  0.5--4 & 1.06--1.13     &  15.3--13\\ \hline
\textbf{P2}            & 1150--1250                  & 425--500      &  3--5     &   5--7      & 2--4      & 4--5 & 1.2--1.3       &  30.5--26\\ \hline
\textbf{P3}            & 2450--2550                  & 710--850        &  49--75    &  75--106   & 21--26      & 22--28 & 2.5--2.6  & 18.3--15.3\\ \hline
\textbf{P4}            & 960--980                 & 850--1000        &  2--3  &   10--13       & 0.6--1.1  &    2.3--2.8 & 0.98--1 & 15.3--13 \\ \hline
\end{tabular}
\end{center}
\end{table}

Microscopic images of the particles indicated that the particles were not perfectly spherical. Therefore, for every single particle used in the experiment we estimate its effective diameter and density using the measurements of its settling velocity through the top and bottom homogeneous density layers (i.e. $V_1$ and $V_2$, respectively) and applying a standard force balance which is described in the next section. The velocity was measured using particle tracking velocimetry system, based on a digital camera (Optronis CL4000CXP, 2304 x 1720 pixels), recording particle positions at suitable frame rates of $60 - 250$ frames per second and processed using the open source software, OpenPTV (\texttt{http://doi.org/10.5281/zenodo.893435}).  

Due to the index-matching, accurate identification of the interface is a non-trivial process. 
We found that the most accurate method to detect the thickness of the interface is to use the particle trajectories of particles P1 and P4 which are most sensitive and travel in the  opposite direction. 
The interface location is inferred from the velocity of particles P1 and P4, released in experiment 2 through the same stratified interface and measured within a single run (not shown here for the sake of brevity). This is achieved by detecting  the position $z$ at which the particle velocity deviates from the constant settling velocity $V_1$ with an uncertainty of $\pm 1.5 $mm. 
The interface position is verified optically, by adding 2 mL of fluorescent dye (Rhodamine 6G mixed in water; sufficiently dilute so it does not affect fluid properties) to the top layer and applying a background illumination. 
The method provides the vertical distribution of intensity which changes from black (0; top layer) to white (1; bottom layer) as shown in  Fig.~\ref{fig:sketch_problem}b. 
For the sake of clarity, we add to Fig.~\ref{fig:sketch_problem}b an overlay of the average intensity vertical profile $I(z)$ (green solid line), and the scaled vertical velocities of P1 (coloured dashed lines for different particles, shown with a quantitative scale in Fig.~\ref{fig:raw_p1p2p3p4}) emphasising the location of the velocity deviations, and the corresponding intensity profile, that define $h$.
The optical interface location is in good agreement with that inferred from  particles P1 and P4.
The density of the top and bottom layers $\rho_1, \rho_2$ respectively, were determined using pycnometer measurements.


\section{Results}\label{sec:results}
\subsection{Interface crossing}\label{sec:settling}

\begin{figure}
        \centering
\includegraphics[width=0.9\textwidth]{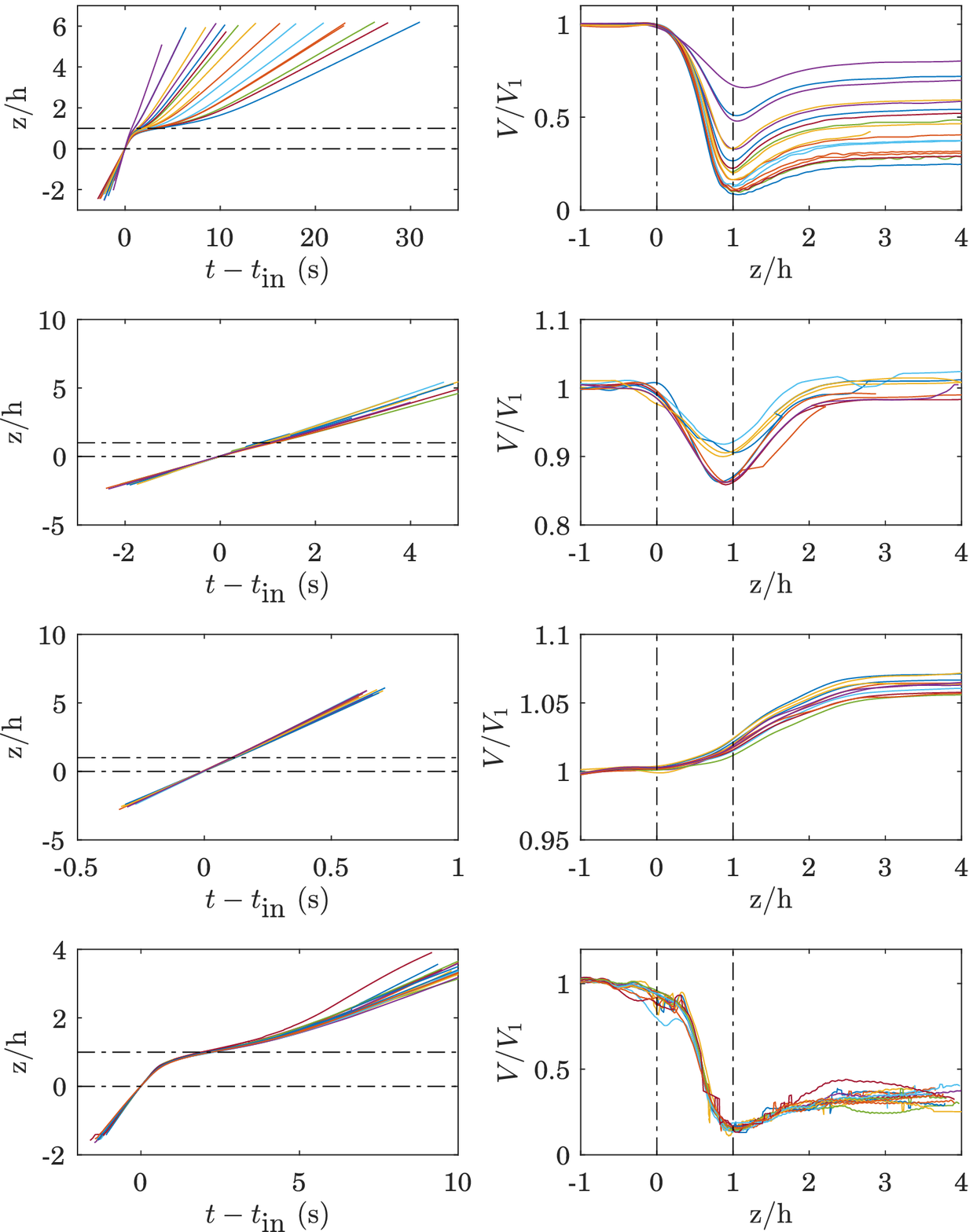} 
                  \caption{(left column) Time trajectories and (right column) normalised velocity $V/V_1$ versus normalised z-position $z/h$, scaled respectively by the time of entering within the stratified region $t-t_\mathrm{in}$ and the entrance of the interface $z/h=0$. (a,b) particles of P1, (c,d) P2 type,(e,f) P3 and (g,h) P4 type. Vertical and horizontal dashed lines indicate the upper and lower boundary of the interface.}    \label{fig:raw_p1p2p3p4}
\end{figure}

This section reports the PTV results for particle types P1-P4, moving between two layers of different density through an interface layer of finite thickness $h/a$. We present the normalised vertical coordinate $z_p(t)/h$ of the individual particles in the left column and normalised vertical velocity $V(z)/V_1$ in the right column of figure~\ref{fig:raw_p1p2p3p4}. The interface position for each experiment is shown by dashed lines at $0$ and $1$.  The time axis is shifted for each particle according to the time the particles enters (from above or from below) into the density interface, entitled $t_\text{in}$. Similarly, the vertical coordinate is defined with respect to the top interface position, $z=0$ for the particles of types P1-P3, and with respect to the bottom interface position, $z=h$ for the particles of type P4. The velocity of the particles is shown as a function of $z_p(t)$ in order to emphasise the change of velocity of the particle as it enters the interface layer $0 < z < h$ and afterwards, when particle leaves the interface layer, i.e. $z > h$. The velocity changes are seen in the trajectories in the left column as the change of slope of $z_p(t)$. 

The behaviour of particles P1 type is shown in figure~\ref{fig:raw_p1p2p3p4}(a-b). In the homogeneous fluid layers, i.e. before and after the passage of the density interface the particles move at constant velocity, $V_1$ or $V_2$. Immediately after entering the interface layer, the particles decelerate considerably.
The velocities of the particles reduce to a minimal value, which is substantially lower than the settling velocity of the particle in either of the layers. After exiting the interface, particles "recover" to the typical settling velocity of the second layer, albeit at an unexpectedly slow rate. Indeed, the recovery is observed for the distances of tens to hundreds of particle diameters. The recovery time was much longer than predicted by Eq.~\ref{eq:beq} with variable $\rho_f(z_p)$, which we attribute to the dynamics of the fluid following the particle from the upper layer and its replacement by the dense fluid in the bottom layer.

The behaviour of P2 type particles is presented in figure~\ref{fig:raw_p1p2p3p4}(c-d). These are relatively small but heavy particles, that have a similar settling velocity in both layers, i.e. $V_1 \approx V_2$, despite the difference of densities $\rho_2 > \rho_1$. This is because $\nu_2<\nu_1$ due to the refractive index matching of the two fluid layers, which negates the effect of the density difference on the settling speed for this particular particle.
For the same reason ($\nu_2 < \nu_1$), the very heavy glass beads of type P3 (see figure~\ref{fig:raw_p1p2p3p4}(e-f)) accelerate. 
The increase of the settling velocity for particles P3 in the bottom layer can be understood by considering the ratio of the settling velocity in both layers that can be approximated as:
\begin{equation}
\frac{V_2}{V_1} = \frac{(1 - \rho_{2}/\rho_p )}{(1 - \rho_{1}/\rho_p)} \frac{\mu_1}{\mu_2}
\end{equation}
\noindent If particle P3 would fall in a non-index-matched fluid, the settling velocity would decrease since $\rho_2 > \rho_1$. However, for the index-matched fluids in the experiment we have that $\mu_2 < \mu_1$, which outweighs the effect of the density difference, thereby causing the particle to -- rather counter-intuitively -- accelerate.
The rising particles of type P4 behave very similar to particles of type P1. Their velocity in figure~\ref{fig:raw_p1p2p3p4}(h) is shown on a negative scale for consistency with our definitions in figure~\ref{fig:sketch_problem}.

\subsection{Determination of $\rho_p$ and $a$}

The equation of motion for a spherical particle in a homogeneous fluid is~\citep[e.g.][]{maxey1983}:
\begin{equation}\label{eq:beq} 
m_p \frac{dV}{dt} = F_{WB} - F_D + F_A + F_H.
\end{equation}
The particle accelerates due to the balance between the
immersed weight ($F_{WB}$), the drag ($F_D$), added mass ($F_A$) and Basset (or history, $F_H$) forces~\citep{maxey1983,Clift2005,Fernando1999}.  Here, the immersed weight $F_{WB}$ and drag force $F_D$ are defined as

\begin{eqnarray}\label{eq:wb}
F_{WB} & = &(\rho_p  - \rho_f)\V_p g \\
 F_D &=& C_D \frac{1}{2}\rho_f |V| V A_p,
\end{eqnarray}
\noindent where $A_p=\pi a^2/4$ is the projected surface area, $\V_p=\pi a^3 /6$ is the particle volume, and $C_D$ is the drag coefficient. The latter can be expressed as a function of the particle Reynolds number~\citep{white1974}:
\begin{equation}
C_D = 0.4 + \frac{24}{Re} + \frac{6}{(1+Re^{1/2})} \label{White}
\end{equation}

The added mass term $F_A$ and the history force $F_H$ in the range of $0<Re<62$ are usually modelled as~\citep[e.g.][and references therein]{odar1964, Fernando1999} 
\begin{eqnarray}\label{eq:F_A}
F_A &=& - C_A \frac{\rho_f}{2} \V_p \frac{dV}{dt} \\ 
F_H &=& -C_H \frac{3d^2}{2}(\pi \rho_f\mu)^{1/2}\int_{-\infty}^t \dot{V}(s)\frac{ds}{(t-s)^{1/2}}
\end{eqnarray}
where $C_A$ and $C_H$ are two empirical coefficients \citep{odar1964}. These coefficients depend on the ratio between the acceleration and velocity square of the sphere $M_A=(dV/dt)/(V^2/a)$ \citep{Fernando1999, Clift2005}:
\begin{eqnarray}\label{Ca_coeff}
C_A &=& 2.1-0.132 M_A^2/(1+0.12M_A^2)\\
C_H &=& 0.48+0.52 M_A^3/(1+M_A)^3
\end{eqnarray}

Although bounds on the particle density and diameter are provided by the manufacturer, these are too large to accurately determine the stratification force.
Thus, we determine the particle density and diameter by making use of the constant settling velocity before and after the interface crossing, for which $dV/dt = F_A = F_H = 0$.
This implies that Eq.~\eqref{eq:beq} simplifies to:
\begin{equation}\label{eq:motion_eq}
(\rho_p  - \rho_f)\V_p g = C_D \frac{1}{2}\rho_f V^{2} A_p 
\end{equation}
For each particle the settling velocities in the two homogeneous layers ($V_1, V_2$) are measured with PTV with an error of $\delta V = \pm 1$ mm/s. These velocities, in conjunction with the drag coefficient given Eq.~\eqref{White}, are substituted into Eq.~\eqref{eq:motion_eq}, using the  initial guess of $\rho_p$ as provided by the manufacturer. A non-linear least squares optimisation (bounded by the range of values provided by the manufacturer) is applied to minimise the error between the measured and estimated velocities. The obtained values of $\rho_p, a$ are within the range provided by the manufacturer, and reported in  tables~\ref{tab:p1_table}--\ref{tab:p4_table}.

\begin{figure}
        \centering
        \includegraphics[width=0.6\textwidth]{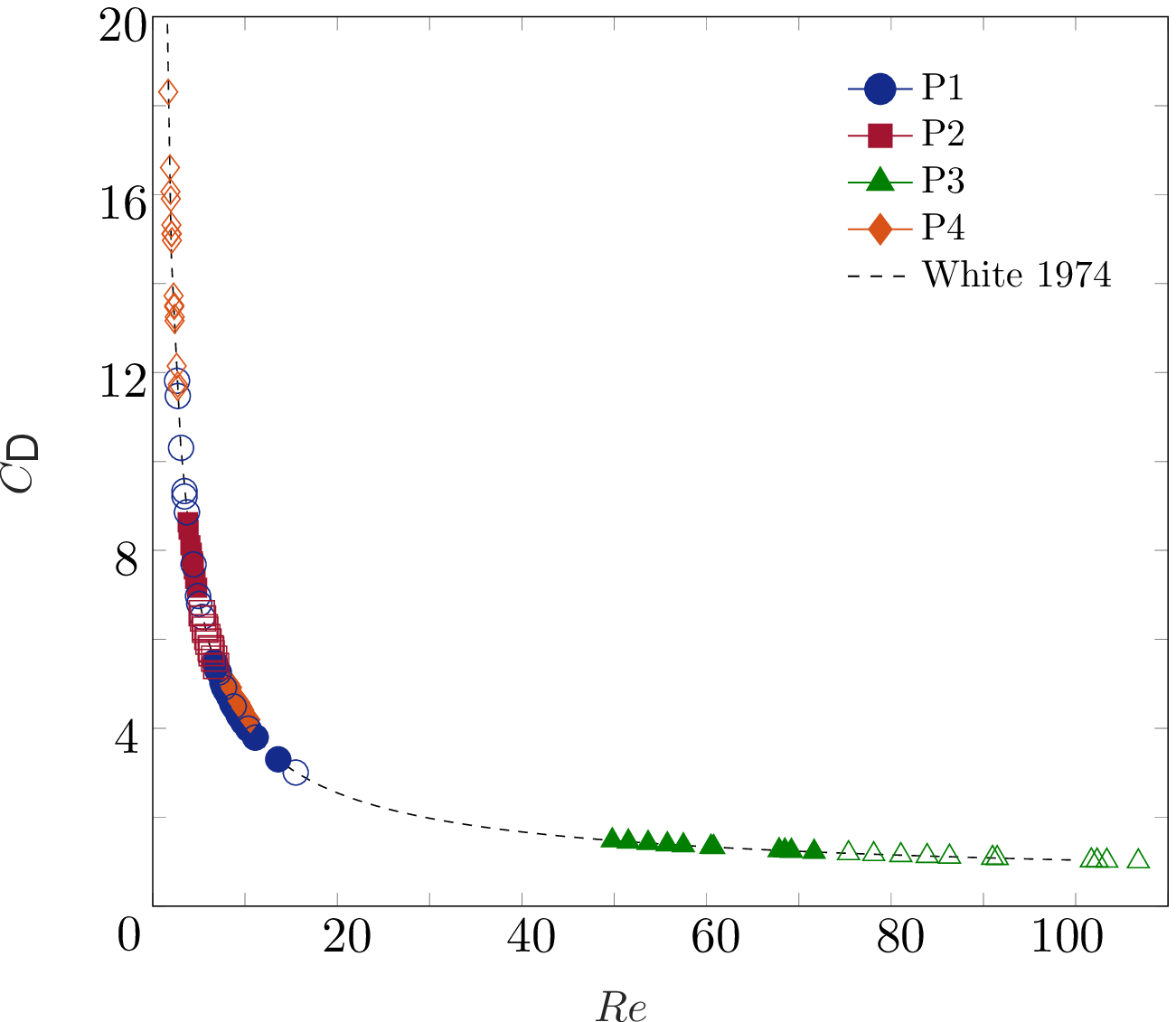}
                  \caption{The dimensionless drag coefficient $C_D = F_D/\frac{1}{2}\rho_f V^{2} A_p$ versus $Re = Va/\nu$. Full markers refer to the particles before the crossing and the  empty markers after the crossing.}
    \label{fig:Cd_Re}
\end{figure}

The drag coefficients of all the particles in both upper and bottom layer are plotted against the particle Reynolds number $Re$ values in figure~\ref{fig:Cd_Re}. The results are in good agreement with \eqref{White}, both before and after crossing the interface (filled vs empty markers).
As the Reynolds number is both a function of the diameter and particle density, the good agreement with  \eqref{White} shows that the constraints on the bounds provided by the manufacturer were not invoked in the constrained optimisation.
Figure~\ref{fig:Cd_Re} demonstrates the relatively large range of particle Reynolds numbers covered in this study.
We note that there is some uncertainty in $V_1$ and $V_2$ for P3 and P4 due to slow variation of the settling velocity (figs.~\ref{fig:raw_p1p2p3p4}(f,h)), but these only very weakly affect the estimates for $\rho_p$ and $a$ as the actual settling speeds deviate from the theoretically constant values by less than 1 percent.

\subsection{Particle motion in the density interface}

The equation of motion Eq.~\eqref{eq:beq} describes the behaviour of a spherical particle in a homogeneous fluid adequately.  However it cannot predict the motion through the interface where the velocity observed to be lower than the expected settling velocity, as shown in figure~\ref{fig:raw_p1p2p3p4} \citep[see also][]{Fernando1999}. SMF also noted that very slow particles, expressed by low $Re$, and also very fast particles at large $Re$, crossed the interface layer at expected settling velocity estimated for the varying density. The authors suggested an additional force in a form of drag and attributed it to the caudal wake appearance.
The wake was filmed and quantified till its rupture at some depth. It was hypothesised that the additional drag exists until some dimensionless depth, $z/a$, inside the interface layer, where the minimal velocity was observed.
A dimensional analysis of the minimal velocity suggested that $V_\text{min}$ is a function of $Re_1, Fr_1$ in the form of:
\begin{equation} \label{fer_scaling}
\frac{V_\mathrm{min}}{(N\nu)^{1/2}} = \alpha Re_1^{n}Fr_1^{m},      
\end{equation}
\noindent where the best collapse of experimental data was found when $m=7/5$, $n=1/2$  and $\alpha = 5.5\times 10^{-2}$ (SMF).

\begin{figure}
        \centering
   \includegraphics[width=1\textwidth]{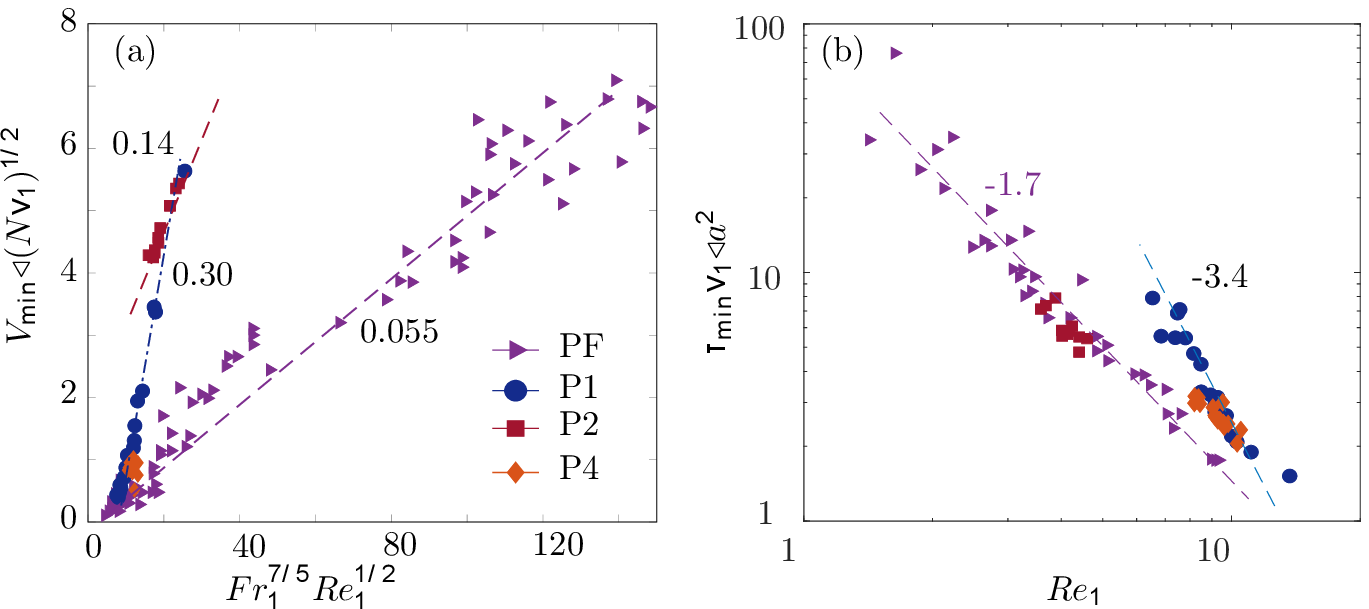}
 \caption{(a) Plot of the normalised $V_\mathrm{min}$ versus $Fr^{7/5} Re^{1/2}$. (b) normalised time to achieve the minimum velocity as a function of $Re_1$. Particles denoted PF have been digitised from \cite{Fernando1999}}
    \label{fig:Vmin_Fr1}
\end{figure}

For the sake of comparison, we plot in figure~\ref{fig:Vmin_Fr1} the results from our measurements for the particles (P1,2,4) together with the data presented in fig.13 in SMF. Note that both in our and SMF results very fast particles do not exhibit a minimal velocity, and are therefore excluded from this figure (in the next section we demonstrate that this type of particles do however experience a stratification force). 
Figure~\ref{fig:Vmin_Fr1}(a) reports the normalised minimum velocity.
The different trends for various particles, emphasised using dashed lines demonstrate that our data does not comply with Eq.~\eqref{fer_scaling}, 
even though some particles in experiments from the parameter range  ($2<Re<15$ and $3<Fr<10$) resemble the particles measured by~SMF and should in principle follow the same trend. 

The reason for these deviations may be explained by dimensional analysis. Indeed, starting from the basic definition of the problem as sketched in figure~\ref{fig:sketch_problem}, one would expect that
\begin{equation}
    V_{\text{min}}=f(\rho_p, \rho_1, N, a, h, V_1, \nu_1, \nu_2),
\end{equation}
where $N$ and $V_1$ were introduced in lieu of $\rho_2$ and $g$, respectively. Using $V_1$, $a$ and $\rho_1$ as characteristic parameters and using the Buckingham-$\Pi$ theorem yields
\begin{equation}
    \frac{V_{\text{min}}}{V_1}=g\left(Re_1, Fr_1, \frac{h}{a}, \frac{\rho_p}{\rho_1}, \frac{\nu_2}{\nu_1}\right),
\end{equation}
implying that this quantity depends on \emph{five} dimensionless quantities.
Thus, even though $Fr_1$ and $Re_1$ might be similar, the other quantities are not and will thus display different behaviour. Here it is noteworthy that the range of $Re_1$ and $Fr_1$ in SMF was obtained by varying the particle diameter of one type with a small variation of density $\rho_p$ and in the same fluids. Therefore neither the effect of the ratio $\rho_p/\rho_1$ nor the effect of the viscosity, $\nu_2/\nu_1$ could be investigated.  

A similar trend is observed for the time it takes particles to attain the minimal velocity ($\tau_\mathrm{min} = t_\mathrm{min} - t_\textrm{in}$). 
SMF found that their data could be characterised by: 

\begin{equation} \label{fer_tmin}
\frac{\tau_\mathrm{min}}{a^2/\nu} = f(Re_1,Fr_1) = \beta Re_1^{l},
\end{equation}
where the best collapse of the data was found when $l=-1.7$ and $\beta = 1.4\times 10^{2}$.
Figure~\ref{fig:Vmin_Fr1}(b) presents the normalised $\tau_\mathrm{min}$ as a function of $Re_1$.
As before, we take the different trends we observe for the different particles as evidence that $\tau_\text{min}$ does not only depend on $Re_1$ and $Fr_1$, but also on the other dimensionless quantities.

\begin{figure}
        \centering
\includegraphics[width=0.7\textwidth]{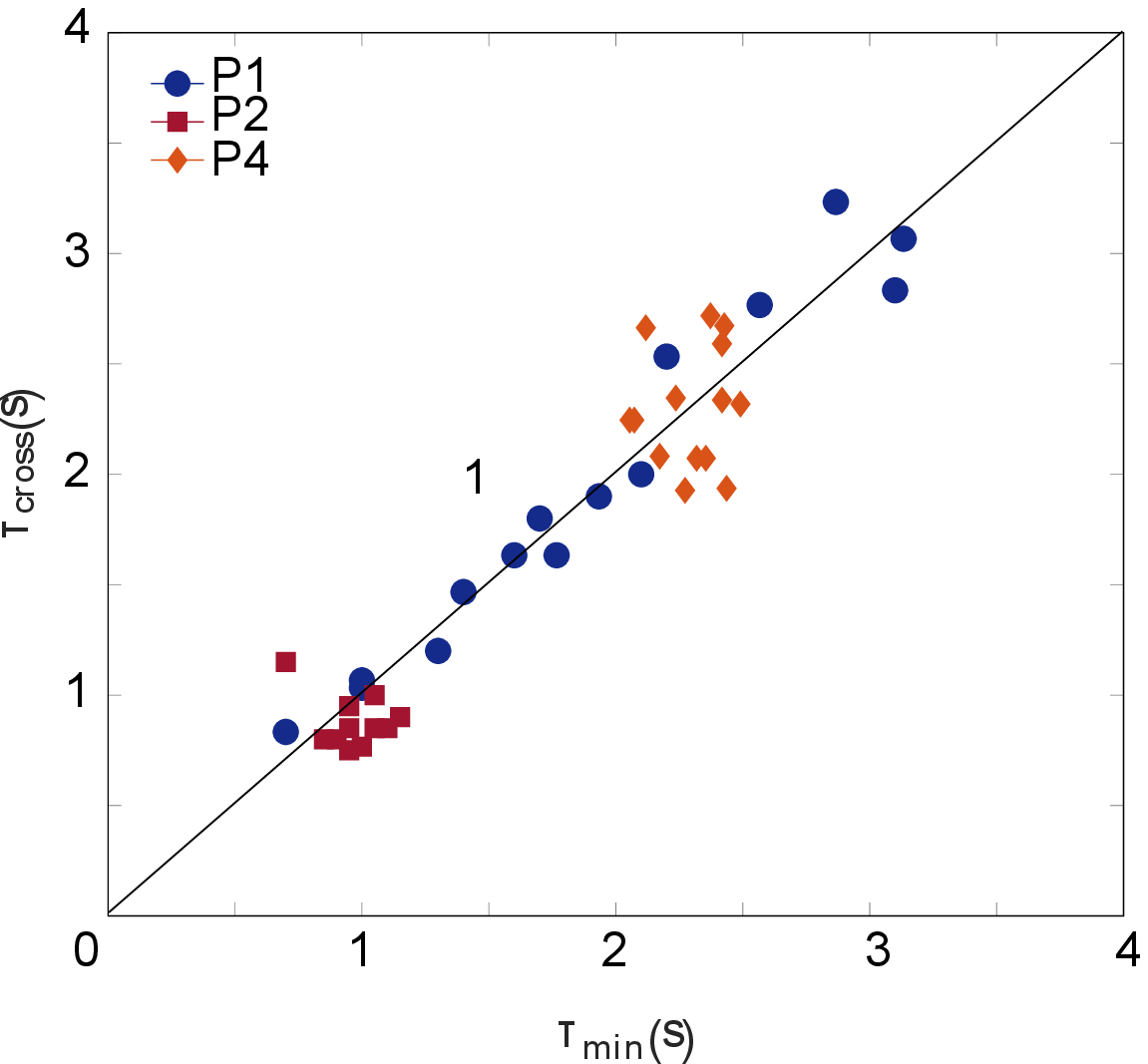}   
                  \caption{$\tau_\text{min}$ versus $\tau_\text{cross}$ for particles P1, P2 and P4 types.}
    \label{fig:tmin_texit}
\end{figure}

After numerous attempts to find a scaling that can collapse all the particle measurements of the minimal velocity value and its time instant, we noticed that the time it takes the particle to cross the interface, $\tau_\text{cross}=t(z_p = h) - t_\text{in}$, is strongly correlated with $\tau_\text{min}$. 
This is shown in  figure~\ref{fig:tmin_texit}. 
Thus, our data indicate that there is no "minimal" velocity inside the interface layer.
This observation is at variance with the assertion that $\tau_\mathrm{min}$ can be attributed to the pinch-off process of the caudal wake which was hypothesised by SMF to cause an instant removal of the stratification force.
It was not possible to explore this observation more directly in this study, because it is not possible to carry out shadowgraphy visualisations of the wake due to the refractive index matching.

\subsection{Determining the stratification force}

\begin{figure}
    \centering
 \includegraphics[width=0.7\textwidth]{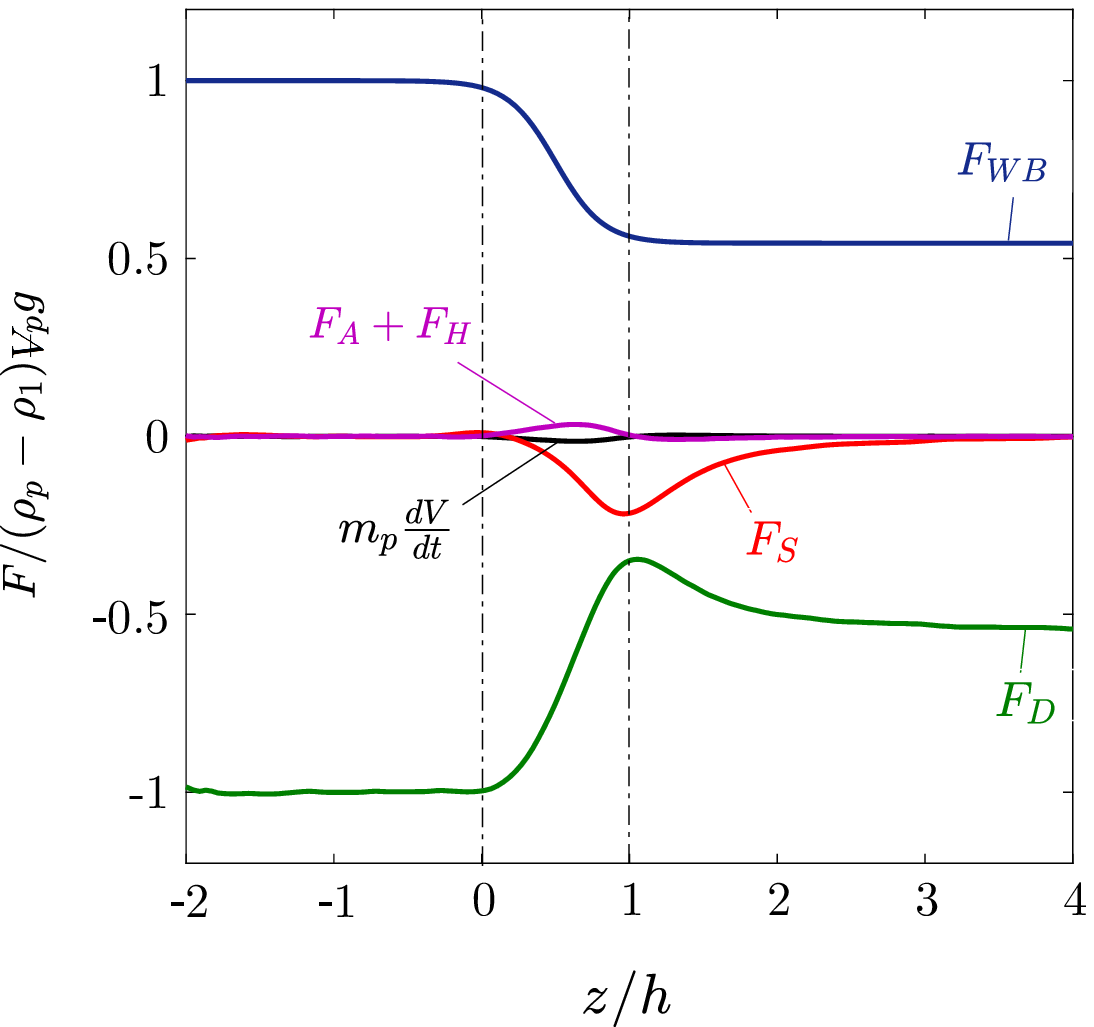}

    \caption{Forces balance on a marked particle P1 type versus $z/h$.
    The immersed weight $F_{WB}$, the drag force $F_D$, the history and added mass forces $(F_A+F_H)$ and the stratification force $F_S$ have been normalised by the immersed weight of the top layer $F_{WB_1} = (\rho_p -\rho_1) \V_p g$. The vertical dashed line are the interface limits.}
    \label{fig:Fs_single}
\end{figure}

\cite{Fernando1999,torres2000,yick2009}, among others, identified that there is a need for an additional force, $F_S$, that accounts for the stratification force due to the distortion of isopycnals:
\begin{equation}\label{eq:eq_stratified}
m_p \frac{dV}{dt} = F_{WB} - F_D  + F_A +F_H - F_S
\end{equation}
The various terms of Eq.~\eqref{eq:eq_stratified}, quantified for a single sample particle (from type P1), using measured particle position, $z_p$ and local density of the fluid $\rho_f(z_p)$, along with its velocity $V_p$ and acceleration $a_p$, are shown in figure~\ref{fig:Fs_single}.

We observe that the Basset and the added mass forces together ($F_A + F_H$) are an order of magnitude smaller than the other force terms. The immersed weight force and the particle acceleration $dV/dt$ term change gradually as a function of the distance from the interface, accordingly with the gradual change of density. As the particle decelerates inside the interface layer, the drag force $F_D$, estimated using the drag coefficient $C_D$ from Eq.~\eqref{White}, decreases. The drag force reaches a minimal value at the exiting edge of the interface and then increases gradually as the particle accelerates to its new steady state value $V_2$. The sum of the forces on the right hand side does not balance the measured particle acceleration and therefore there is an additional force term, $F_S$ that for this particular particle reaches approximately half of the drag force value. 
We note that the stratification force increases approximately linearly as the particle moves into the interface stratified layer, reaches a maximum value in proximity of $h$,  and after the crossing, the force magnitude decreases (apparently exponentially) with a certain recovery time scale.  During this time the particle velocity grows gradually from its minimal value to the settling speed of the bottom layer (or upper layer for particles of type P4) as was observed in figure~\ref{fig:raw_p1p2p3p4}.

\begin{figure}
    \centering
 \includegraphics[width=0.7\textwidth]{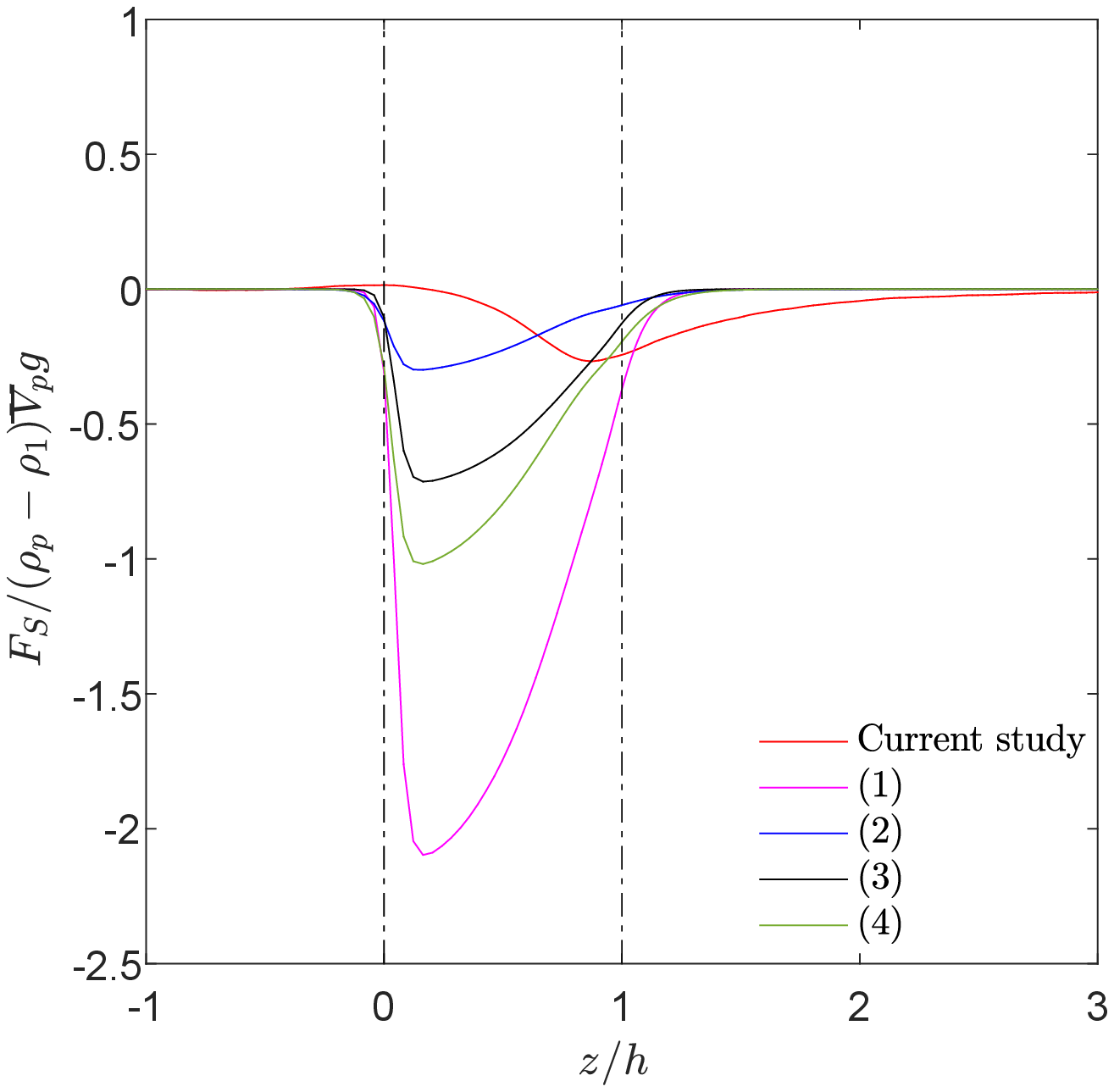}

    \caption{Measured stratification force on a marked particle P1 type versus $z/h$ and application of drag enhanced models valid in a linearly stratified fluid summarised in Table~\ref{tab:cd_models}.
    The forces $F_S$ have been normalised by the immersed weight of the top layer $F_{WB_1} = (\rho_p -\rho_1) \V_p g$. The vertical dashed line are the interface limits. }
    \label{fig:Fs_single_modelcomp}
\end{figure}        
     
Before presenting the drag force for all experiments, we demonstrate that none of the existing drag formulations are able to represent the inherently transient behaviour of the stratification force $F_S$.
Figure \ref{fig:Fs_single_modelcomp} shows a comparison between the measured $F_S$ and the parameterisations by \cite{yick2009}, \cite{Candelier2014}, \cite{Doostmohammadi2014} and \cite{zvirin1975}, which are reported in table \ref{tab:cd_models}.
As is clear from the figures, all models are unable to capture the transient behaviour of $F_S$ as it enters the interface layer. 
This is not surprising -- the models were obtained for linearly stratified fluids for which such transients do not exist. In addition, the models were developed for particles for which $Re$ was much smaller than those reported here.
We also implemented the perturbation force proposed by \cite{Candelier2014}, which introduce a time delay due to the transient behaviour of the memory force. The resulting $F_S$ is not reported since the force has been estimated being orders of magnitude smaller than measured net $F_S$.\\

\begin{table}
\begin{center}
\def~{\hphantom{0}}
\caption{Drag enhanced models according to (1) \cite{yick2009}, (2) \cite{Candelier2014}, (3) \cite{Doostmohammadi2014}, (4) \cite{zvirin1975}.\label{tab:cd_models}}
\begin{tabular}{|c|c|c|c|c|} 
\hline
\textbf{Model} & (1) & (2)  & (3) & (4) \\
\hline
\textbf{$C_{DS}=$}    &  $C_D(1+1.9 Ri^{1/2})$       &   $C_D(1+0.3 Ri^{1/4})$       & $0.67 C_D(Ri^{1/2})$   &  $C_D(1+ Ri^{1/3})$   \\ \hline
\end{tabular}
\end{center}
\end{table}

\begin{figure}
        \centering
\includegraphics[width=1\textwidth]{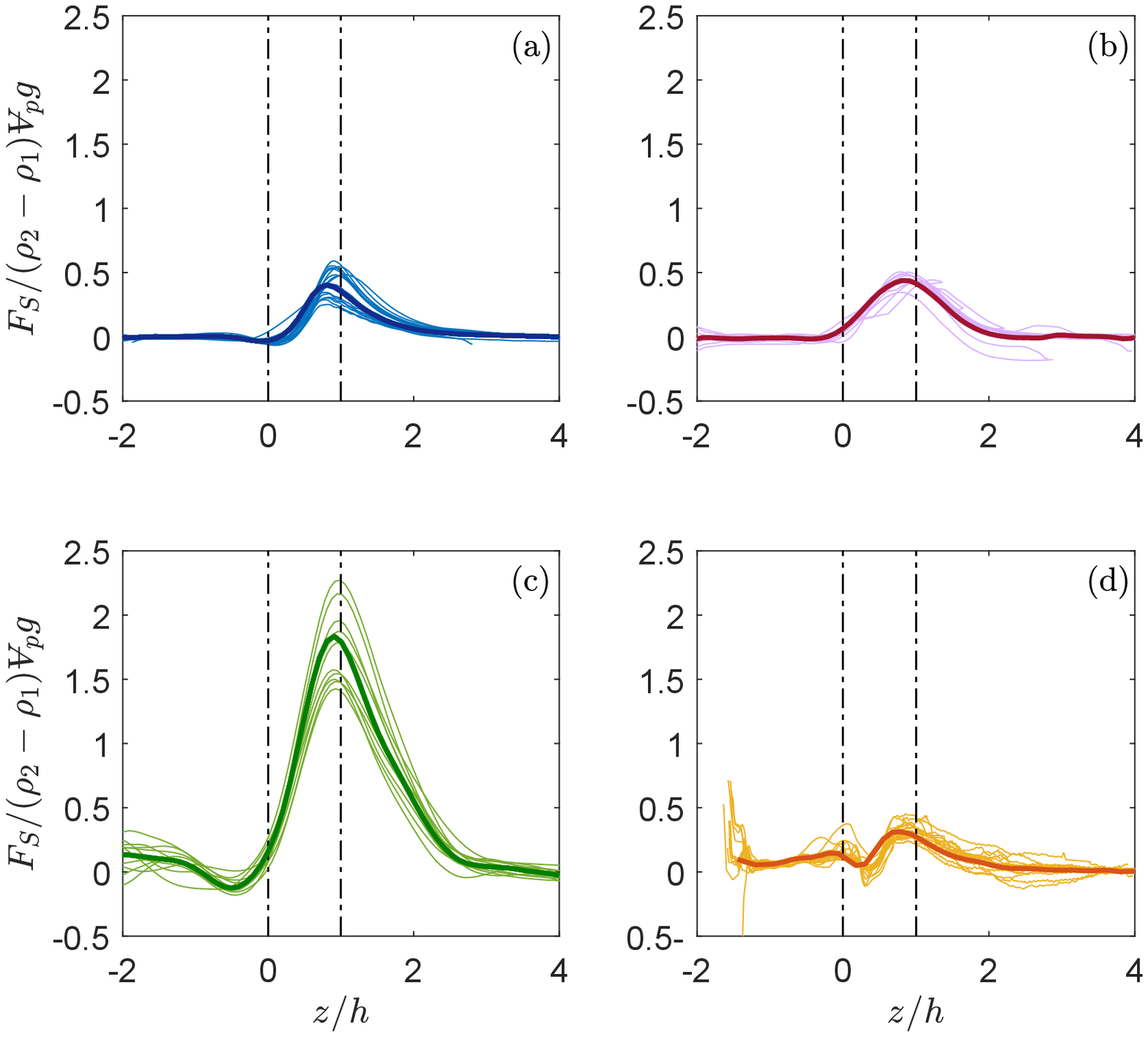}
                  \caption{Normalised stratification force $F_S$ for particle types (a) P1, (b) P2, (c) P3, and (d) P4.  Each lighter curve is
an independent experimental measurement, while the average trend for each type of particle is emphasised as a bold line.}
    \label{fig:Fs_all}
\end{figure}

Figure~\ref{fig:Fs_all}a-d shows the stratification force $F_S$, normalised with the immersed weight, presented versus $z/h$ for all particles types. The vertical dashed lines represent the edges of the density interface. In contrast with~SMF, we observe that the magnitude of the force $F_S$ gradually increases for heavy and fast particles belonging to the type P3. The particles enter the interfacial layer at relatively high Reynolds number ($Re_1$), but clearly experience a stratification force $F_S$. The very short crossing time leads to very subtle change of particles velocity within the interface. Nevertheless, the effect of stratification is clearly evident also after particles exit the interface layer and move into the bottom homogeneous density layer. 

\section{Model of the force $F_S$ and its properties}
\label{sec:model}

\subsection{Model development}

As demonstrated in the previous section, none of the standard parameterisations were able to reproduce the time-dependence of the stratification force $F_S$ appropriately.
We thus develop a simple phenomenological theory that models 
$F_S$ and subsequent particle motion as it moves through the interface layer. The model is able to accurately predict the motion of particles of all types in our experiments, and, in addition, particles from the SMF data, as shown in \S\ref{sec:fernando}.
It serves both as a useful model that captures the dynamics of the particle paths in this study and in that of SMF and it is envisioned that it can be replaced by a more sophisticated model in the future (for example that parameterises the memory force).
The core of the model is based on a parameterisation of the stratification force $F_S(t)$ of the form:
\begin{equation}\label{eq:FS_model}
F_{S} = (\rho_1 - \rho_f) \V_c(t) g
\end{equation}
Here, $\V_\text{c}$ is an effective fluid volume attached to the particle. 
We emphasise that $\V_c$ should not necessarily be interpreted as the caudal wake observed by SMF that changes volume as particle penetrates deeper into an interface layer and ruptures at some finite depth.
The main purpose of $\V_c$ is to provide a prediction for $F_S$. It can be interpreted in several ways, including as a time-dependent wake density.

\begin{figure}
    \centering
    \includegraphics[width=0.8\textwidth]{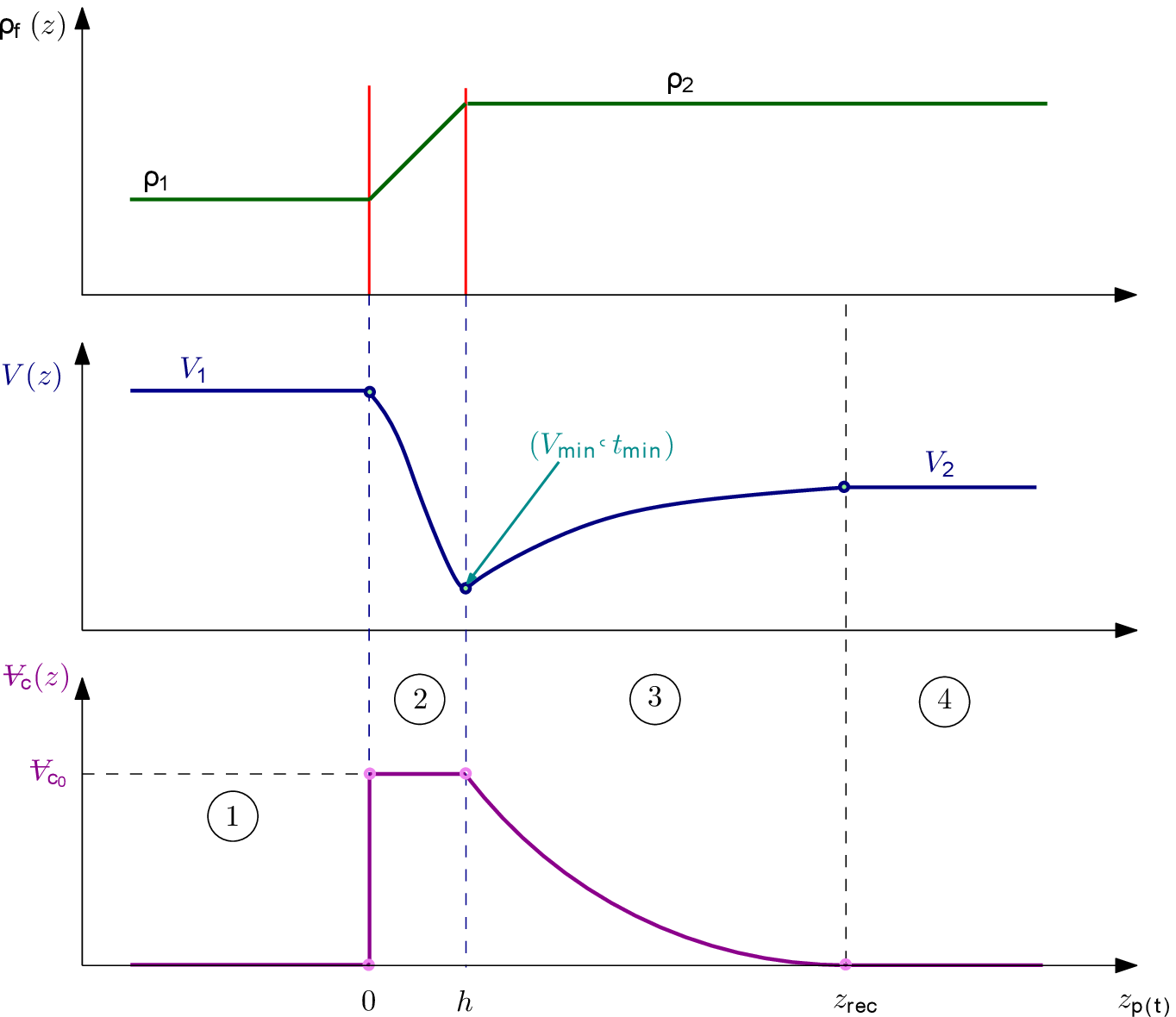}
    \caption{Sketch of the proposed model for the solution of the two layers problem. The density of the surrounding fluid at the $z_p$, the particle vertical velocity, the proposed concept of the caudal volume. In the following we will address the time measured from the entrance moment, i.e. $\tau_\text{min} = t_\text{min} - t (z=0)$}. 
    \label{fig:model}
\end{figure}

The effective volume $\V_c(t)$ is modelled as
\begin{equation}\label{eq:v_c}
\V_c(t)=  \left\{
\begin{array}{ll}
         \V_{c0}   &       0 < z_p(t) \le h \\
         & \\
         \V_{c0}    \exp\left(- \displaystyle \frac{t-t_h}{\tau_\text{rec}}\right)  &      z_p(t) > h \\
\end{array} 
\right. 
\end{equation}
\noindent where $\V_{c0}$ is the base or initial volume, $t_h$ is the time when the particle leaves the interface layer and moves into the second homogeneous density layer, and $\tau_\text{rec}$ is the timescale over which the particle reaches its new terminal velocity. This parameterisation is best explained using the sketch in figure~\ref{fig:model}, which shows the vertical profiles of density $\rho_f$ (top layer is on the left, bottom layer is on the right and the interface is approximated as a linear stratification, for simplicity), particle velocity, $V(z_p(t))$ and the volume $\V_\text{c}$. We propose a simple mechanism that can be explained a sequence of events according to the four regions, marked on the figure: 
\begin{enumerate}[(1) ]
\item Particle settles (or rises) in a homogeneous layer at constant terminal settling velocity $V_1$ defined by the standard equation of motion and the balance of the immersed weight with the viscous drag force.
\item Particle enters the interface layer  - we assume that a small volume $\V_{\text{c}0}$ of a light fluid ($\rho_1$), (or a heavy fluid for the rising particle) remains attached to the particle as it moves into the interface layer. 
For density interfaces of thickness $h/a \sim \mathcal{O}(10)$ as in our experiment, this volume is assumed to remain constant during the time taken to cross the interface. As the initial volume does not dilute, this results in a stratification force $F_S$ which increases in strength, thus leading to a deceleration of the particle proportional to the slope of the density profile.
\item  Once the particle enters the new layer, $F_S$ reduces approximately exponentially over a timescale $\tau_\text{rec}$, which we model via $\V_\text{c}$.
\item Particle reaches its terminal velocity when $F_S \xrightarrow{} 0$. 
\end{enumerate}

In order to develop a predictive model, it is necessary to parameterise  $\V_{c0}$ and $\tau_\mathrm{rec}$ in Eq.~\eqref{eq:v_c} using the experimental data.
Note that these data also include the results of the very fast particles P3 that do not exhibit a minimal velocity but do experience a discernible stratification force $F_S$.

\begin{figure}
\centering
\includegraphics[width=0.45\textwidth]{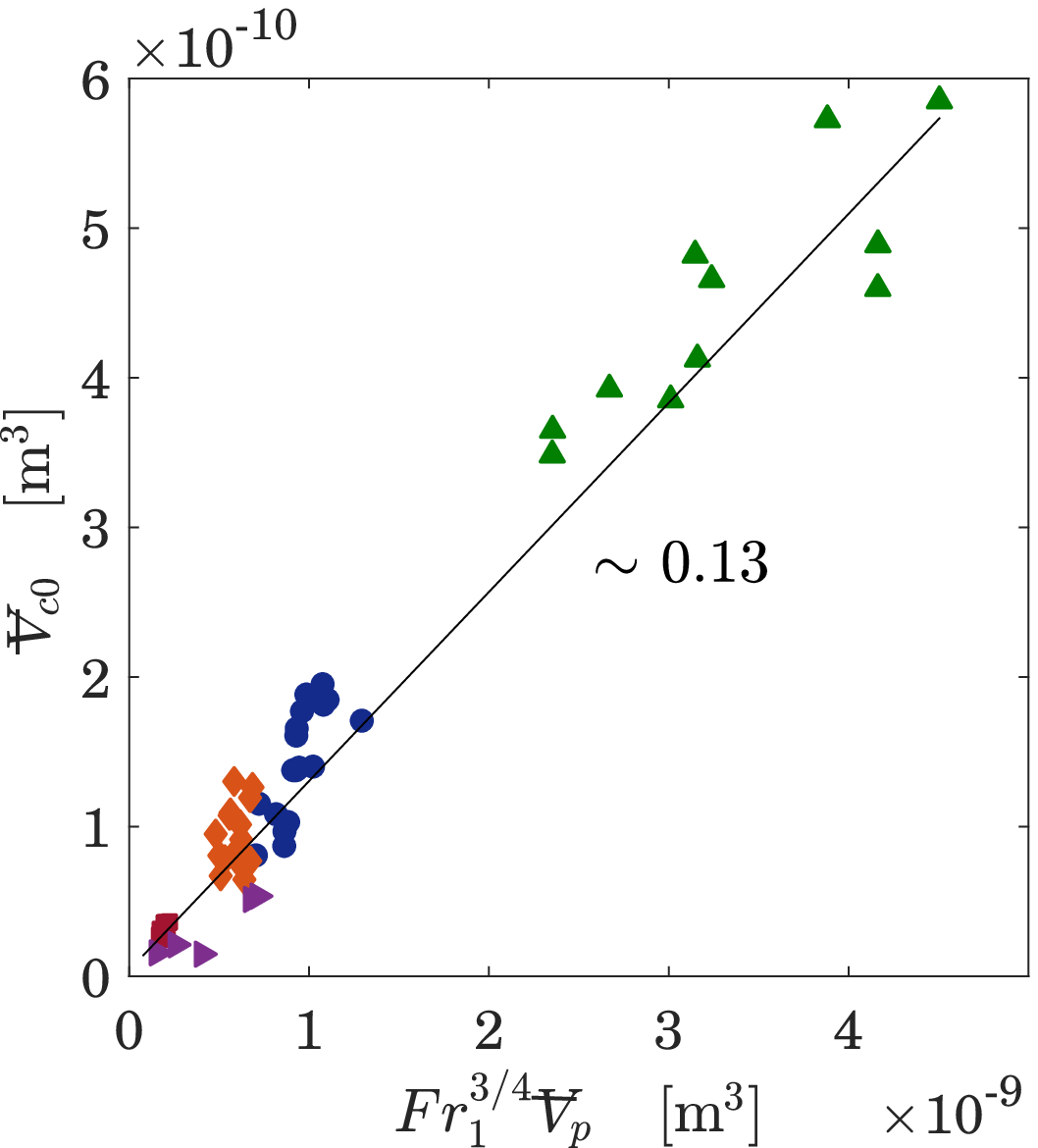}
~
\includegraphics[width=0.45\textwidth]{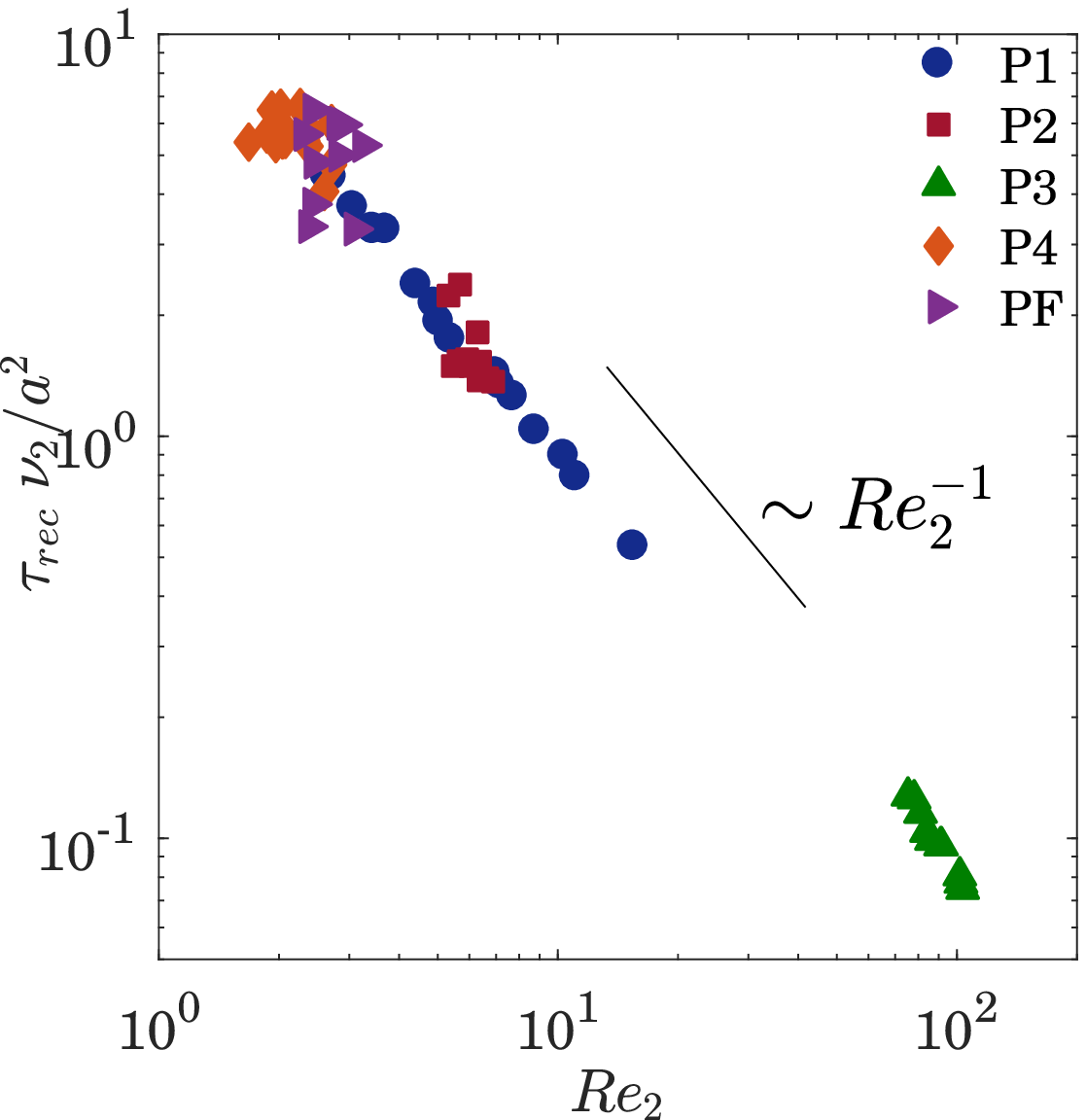}
\caption{(a) Caudal volume $\V_{c0}$ vs $Fr_1^{3/4} \V_p$; (b) normalised recovery time $\tau_{\mathrm{rec}} \nu_2/a^2$ versus $Re_2$.\label{fig:model_params}}
\end{figure}   

The recovery time $\tau_\textrm{rec}$ and the caudal volume were determined by minimisation of the difference between the simulated (Eq.~\ref{eq:FS_model}) and experimental stratification force (fig.~\ref{fig:Fs_all}).
The values of $\V_{c0}$ are reported in fig.\ref{fig:model_params}(a)
versus the "entrance" Froude number $Fr_1$, or the ratio of two times scales - the buoyancy time scale $t_{N} = 1/N$ and the characteristic time of a particle $t_\mathrm{p_1} = a/V_1$.  
The simplest best fit data in the range $2 < Fr_1 < 28$ results in:
\begin{equation}
\V_{c0} \approx 0.13 Fr_1^{3/4} \V_p
\label{eq:V_eff}
\end{equation}
 This parameterisation works well for our experimental data, although care should naturally be taken if used in a different part of the parameter space. 
Particles with $t_\mathrm{p_1} \ll t_{N}$ according to this parameterisation, will have longer time to experience the surrounding density changes. Particles with  $t_\mathrm{p_1} \gg t_N$ ($Fr_1<1$), will cross the interface in a quasi-steady motion.
 These particles effectively move too slow to drag the additional fluid, as the surrounding fluid can return to the original position without distorting the isopycnals during the particle motion.
 Here we note that this observation differs substantially from the predictions in \cite{Zhang2019}. For the cases under consideration, $Re\gg 1$ and $Fr \gg 1$, which implies they are in regime R3. In this regime, the stratification force is dominated by the vorticity field, which is expected to scale as $F_S\sim Fr^{-1} Re^{-1/2}$. However, our results indicate that $F_S \sim Fr^{3/4}$, but this is not surprising since the work in \cite{Zhang2019} deals with linearly  stratified environments in which the wake will respond slowly to changes in the environment, whereas the case under consideration here features very rapid changes in the environment which will produce a significant time-dependent response.

 The recovery time scale $\tau_\mathrm{rec}$, normalised by the
 viscous time scale $t_{\nu_2}=a^2/\nu_2$, is shown in figure~\ref{fig:model_params}(b). The parameter is plotted versus the Reynolds number in the bottom layer (using terminal velocity $V_2$) $Re_2$ and its best fit suggests:
\begin{equation}
\frac{\tau_\mathrm{rec}\nu_2}{a^2} \approx 13  \, Re_2^{-1}\label{eq:tau_rec}
\end{equation}

This relation implies that the timescale over which the particle recovers is associated with the advection time scale in the second layer as $\tau_\mathrm{rec} \sim a^2/(\nu_2 Re_2) = a/V_2$.

\subsection{Model performance}\label{sec:performance}

The particle model is represented by two simultaneous equations
\begin{eqnarray}
  \label{eq:model1}
  \frac{\d z_p}{\d t} & = & V \\
  \rho_p \V_p \frac{\d V}{\d t} & = & (\rho_p-\rho_f) \V_p g - \frac{1}{2} \rho C_d(Re) A_p |V| V - (\rho_f-\rho_1) \V_{c} g 
  \label{eq:model2}
\end{eqnarray}
where the Basset force $F_B$ and added mass force $F_A$ have been neglected in Eq.~\eqref{eq:model2}. Both the base caudal volume $\V_{c0}$ and recovery time $\tau_\text{rec}$ are provided by the empirical relations \eqref{eq:V_eff} and \eqref{eq:tau_rec}, respectively.

\begin{figure}
     \centering
 \includegraphics[width=1\textwidth]{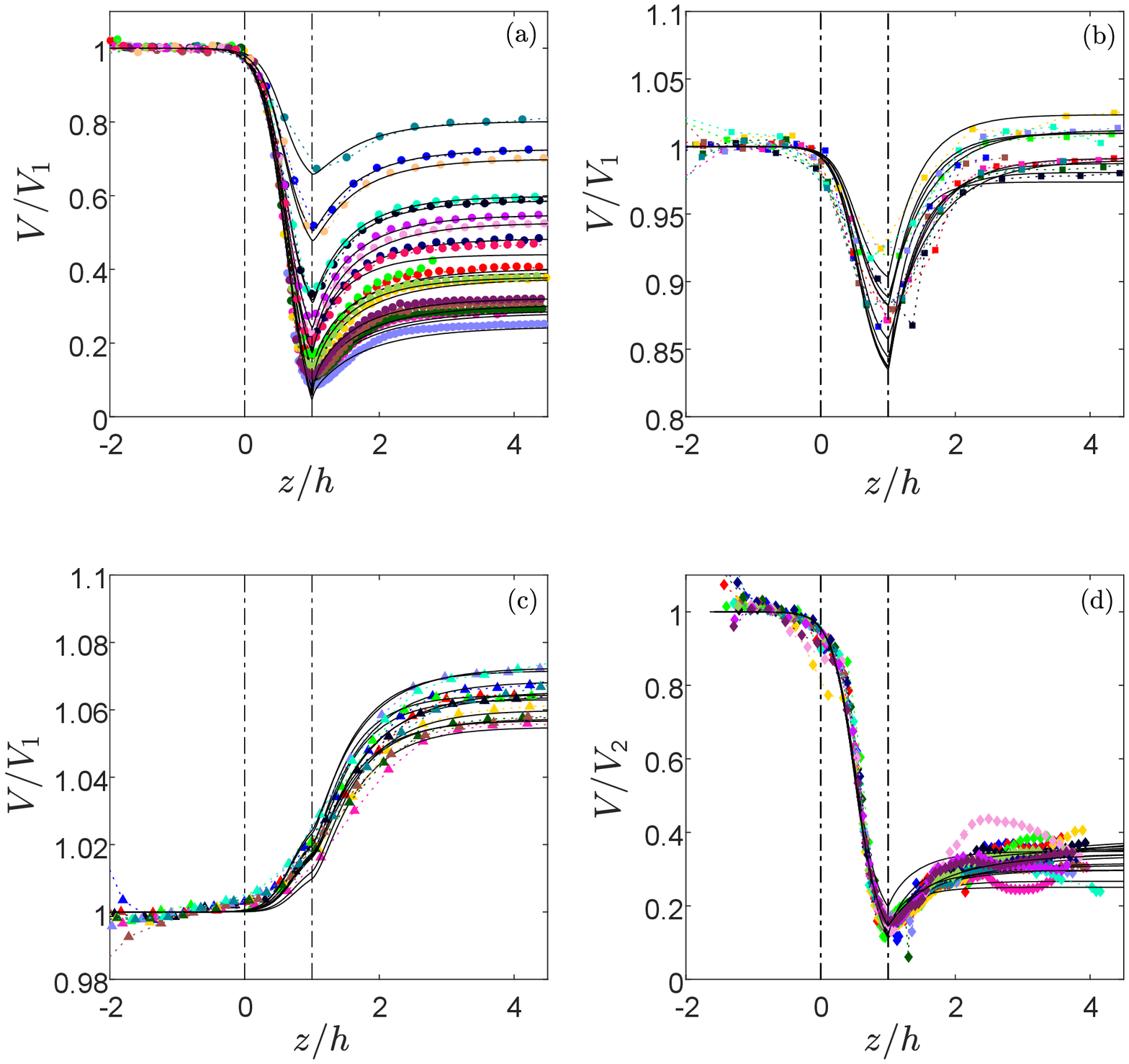}
     \caption{Velocity profiles normalised by the settling velocity $V_i$ ($i=1$ for particles P1,P2,P3 and $i=2$ for particle P4) as function of the normalised distance from the interface $z/h$ for particle types (a) P1, (b) P2, (c) P3, and (d)P4. The lines are the numerical simulations obtained by integrating in time Eq.\eqref{eq:eq_stratified}.}
     \label{fig:UniqModel_results}
 \end{figure}

The example is shown in figure~\ref{fig:UniqModel_results}(a-d) for the particles from the distinct groups, including large and slow particles P1, small and dense particles P2, very fast and heavy glass beads P3 and buoyant (rising) light particles P4. Clearly, the proposed model shown as solid curves, predicts accurately the motion of all these types of particles (symbols) in terms of their velocity $V_p$. The velocity is normalised for the sake of comparison by their initial velocity $V_1$ for P1,2,3 and $V_2$ for the rising particle. Note that the solid curves are discontinuous in the gradient at $z/h = 1$ because of the simple piece-wise model \eqref{eq:v_c}. 

\subsection{Comparison with data set from SMF}\label{sec:fernando}

\begin{figure}
    \includegraphics[width=1\textwidth]{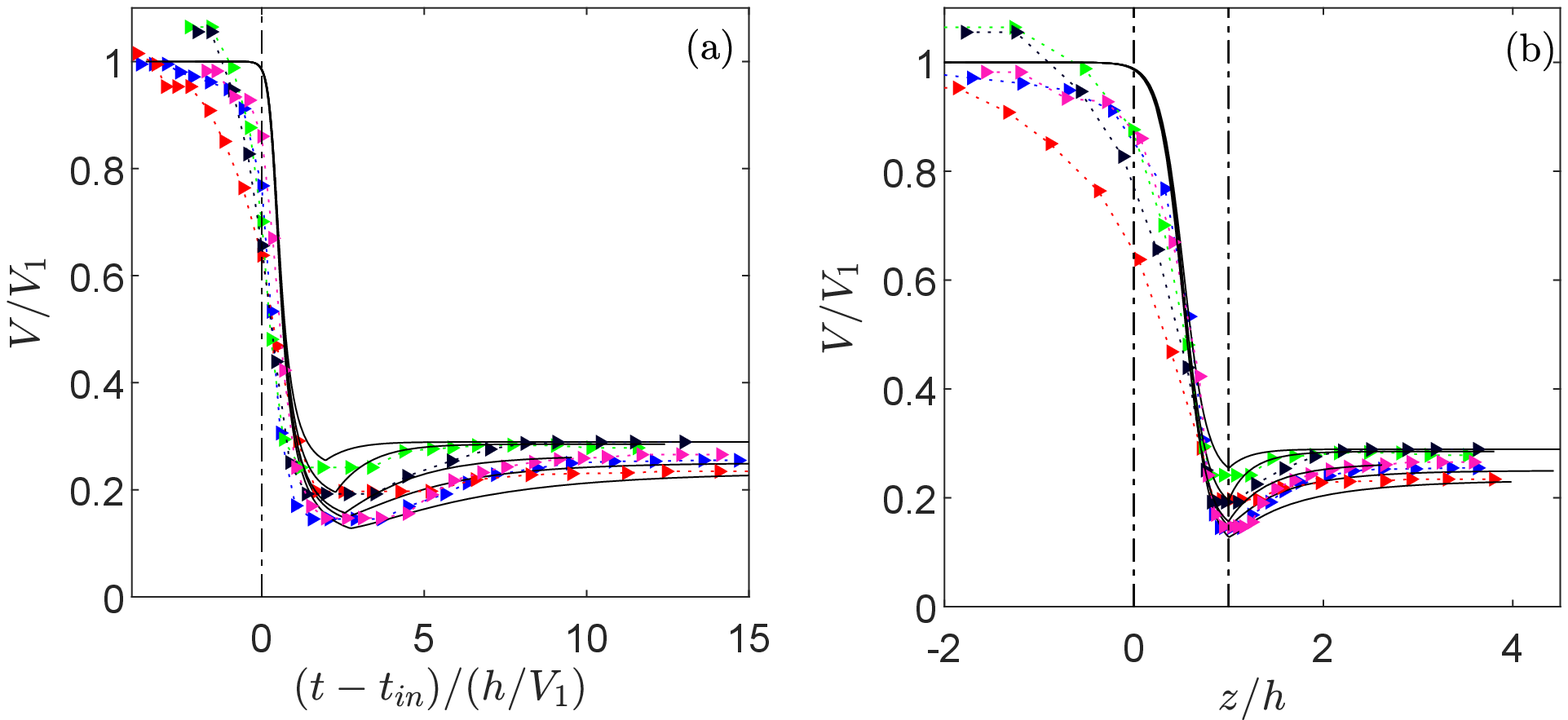}
\caption{Normalised velocity reported versus normalised time (a). Normalised particle velocity versus the normalised distance from the interface $z/h$ (b). The numerical solution (continuous lines) are overlapped to the experimental data from SMF (markers) for few sample particles reported in table~\ref{tab:Fernando_table}.}
    \label{fig:fig_Fernando}
\end{figure}

We apply the same model to a representative set of trajectories digitised from SMF in figure~\ref{fig:fig_Fernando}. 
First we had to use the trajectories to reconstruct the size and density of each particle  ($a$, $\rho_p$), using the constant terminal velocity in the homogeneous layers, estimated from the figure of~SMF. 
We reconstructed the values of $\rho_p$ and $a$ of the digitised data  from the figures of~SMF with a accuracy of $\delta \rho_p = \pm1$ g/cc and $\delta a=\pm10 \mu m$.
Using the reported fluid properties ($\rho_1,\rho_2,\nu_1,\nu_2,h$), estimated settling velocities ($V_1,V_2$), the reconstructed particles properties ($a$, $\rho_p$), and the values of the parameters ($\V_{c}$,$\tau_\mathrm{rec}$) according to our model, we can simulate the trajectories of settling particles.
In figure~\ref{fig:fig_Fernando} we present the measured vertical velocity as symbols and the modelled trajectories as solid lines as a function of the time (figure~\ref{fig:fig_Fernando}(a)) and as a function of the distance from the interface (figure~\ref{fig:fig_Fernando}(b)). Despite the small mismatches around the entrance to the presumable interface position, the model (using only particle and fluid parameters) can predict reliably the motion of the particles in a different parameter range, with most important features such as a position and value of a minimal velocity and the recovery time to the settling velocity in the bottom layer. 

\subsection{Model predictions and analysis}

Next, we explore the properties of the proposed model to predict various phenomena observed in the literature and in our experiments. To keep the analysis tractable, it will be assumed that the kinematic viscosity is constant. 

\subsubsection{Penetration Froude number}
Not the entire ($Fr_1$, $\rho_p/\rho_1$) phase space is realisable due to the physical requirement that  $\rho_p > \rho_2  > \rho_1$ which is needed to ensure that the particle falls through the layer (it can be applied also for the light particles for which $\rho_p < \rho_1 < \rho_2$). The limit at which the particle will not penetrate into layer 2 occurs when $\rho_p = \rho_2$. Substitution into the definition of $Fr_1^2= V_1^2 / N^2 a^2$, using the settling velocity $V_1$ \ \eqref{eq:motion_eq} and the definition of $N^2$ \eqref{eq:N} results in 
\begin{equation}
  Fr_{1;\text{pen}}^2 = \frac{2}{3 C_d(Re_1)} \frac{h}{a} \left(1+\frac{\rho_p}{\rho_1}\right)
  \label{eq:Frpen}
\end{equation}

\subsubsection{Levitation Froude number}
Even for a particle with $Fr_1 > Fr_{1;\text{pen}}$, the  stratification force $F_S$ can cause the particle to stop in the density interface. This is the physical mechanism causing particle levitation, as described in \cite{Adalsteinsson2004}.
In reality, the particle will only levitate temporarily until its wake detaches after which it can continue its journey.
However, this aspect of the physics is not represented in our model, and therefore the particle will not cross the interface.
The limit case that can be used to infer the limit Froude number below which levitation may occur, $Fr_{1;\text{lev}}$, is to assume that the particle comes to rest right at the end of the density interface, which suggests a force balance of the form
\begin{equation}\label{eq:rhovg}
(\rho_p - \rho_2) \V_p g = (\rho_2 - \rho_1) \V_{c0} g.
\end{equation}
Elimination of $\rho_2$ can be achieved using the definition for $N^2$. Using the definition $Fr_1^2$ and the definition of the steady state velocity Eq.~\eqref{eq:motion_eq}, we obtain an implicit equation for $Fr_{1;lev}$:
{
\begin{equation}
\frac{\rho_p}{\rho_1} = \frac{3}{2} C_d(Re_1) \frac{a}{h} Fr_{1;\text{lev}}^2 - 0.26 Fr_{1;\text{lev}}^{3/4} - 1.
 \label{eq:Frlev}
 \end{equation}}
Here we substituted \eqref{eq:V_eff} to eliminate $\V_{c0}/\V_p$. 
Levitation may occur when $Fr_{1;\text{pen}} < Fr_1 < Fr_{1;\text{lev}}$; this will be denoted the levitation regime. 
Levitation is predicted for the entire parameter space ($Re_1$, $Fr_1$, $h/a$, $\rho_p/\rho_1$). 
This can be verified by requiring that $Fr_{1;\text{lev}} \ge Fr_{1; \text{pen}}$. Using \eqref{eq:Frpen}, \eqref{eq:Frlev}, it follows that the criterion for the possibility of levitation is $C_d>0$ which is valid under all circumstances.

\subsubsection{Predictions of $V_\mathrm{min}$ and $\tau_\mathrm{min}$}

\begin{figure}
\centering
\includegraphics[width=.9\textwidth]{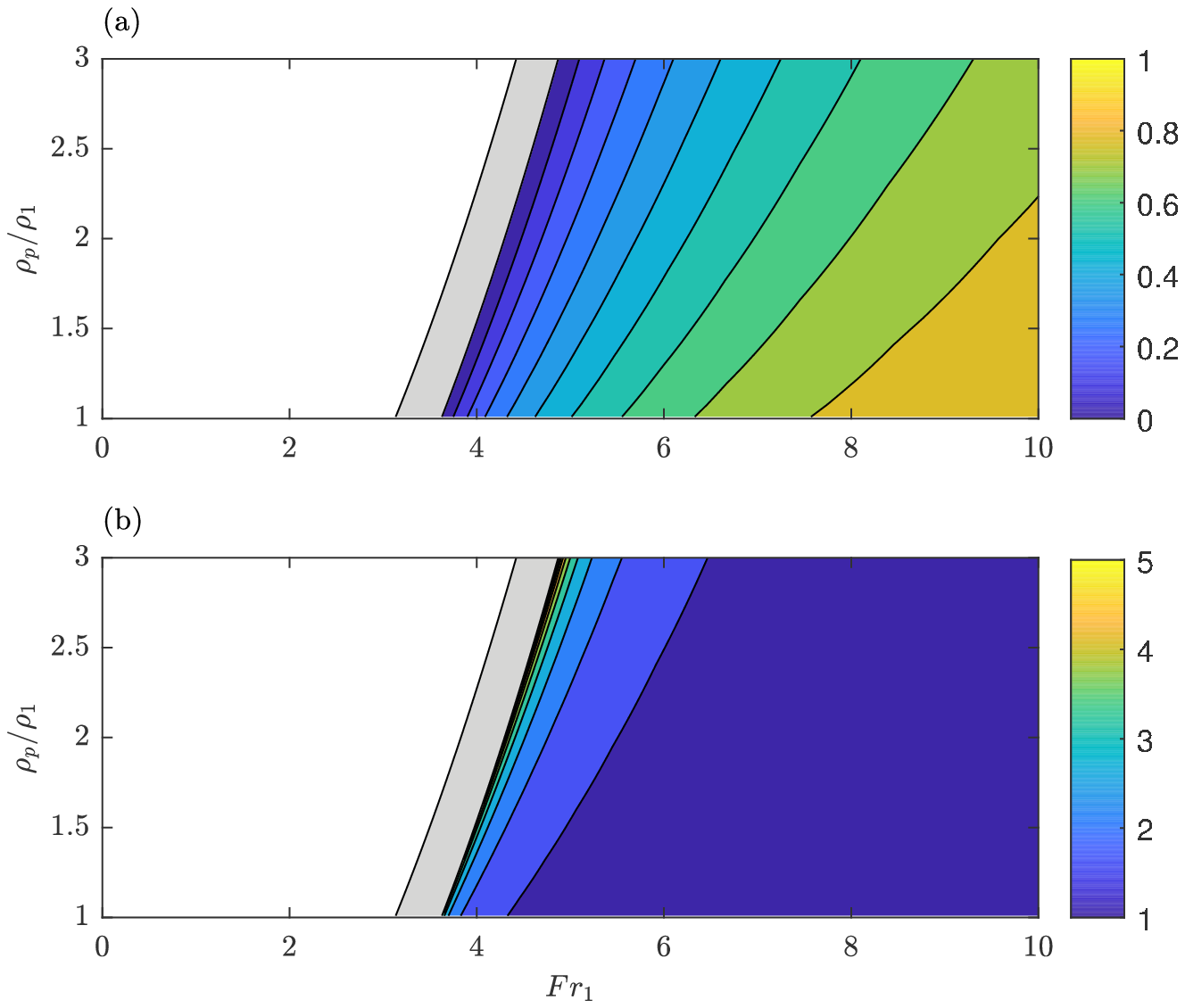}
\caption{($Fr_1$, $\rho_p/\rho_1$) parameter space plots at $Re_1=10$ and $h/a=30$ of: a) minimum velocity $V_\text{min} / V_1$ and b) time to minimum $\tau_\text{min} V_1/h$ (= dimensionless crossing time). The white area is physically unrealisable ($Fr_1 < Fr_{1;\text{pen}}$), and the grey area is the levitation regime ($Fr_{1;\text{pen}} < Fr_1 < Fr_{1;\text{lev}}$).}
\label{fig:parameterspace}
\end{figure}

Here we will explore the dependence of $V_\text{min}$ and $\tau_\text{min}$ on $Fr_1$ and $\rho_p/\rho_1$. We note that the subscript might be misleading and remind the reader that in our experiments both values coincide with the crossing time and and the related velocity of the particle. The dependence on $Fr_1$ and $\rho_p/\rho_1$ of the minimum velocity $V_\text{min}$ and time to minimum $\tau_\text{min}$ for a particle at fixed $Re_1=10$ and $h/a=30$ was calculated by running a series of simulations using Eqs.~(\ref{eq:model1}-\ref{eq:model2}). 
The minimum velocity $V_\text{min}$, scaled by the entrance velocity $V_1$, is shown in figure~\ref{fig:parameterspace}(a). The white area in  this figure denotes the physically inaccessible area for which $Fr_1 < Fr_{1;\text{pen}}$.
The grey area denotes the levitation regime for which $Fr_{1;\text{pen}} < Fr_1 < Fr_{1;\text{lev}}$. The model is not suitable to operate in this regime.
Close to the levitation Froude number $Fr_{1;\text{lev}}$, the particles come to a practical stand-still. 
As $Fr_1$ increases, $V_\text{min}/V_1$ becomes larger until it approaches the limit value 1, in which the particle traverses the density interface undisturbed.
This makes sense as $Fr_1 \rightarrow \infty$ implies that $N^2 \rightarrow 0$, which in turn implies that the stratification strength reduces to zero.

Figure~\ref{fig:parameterspace}(b), which shows the minimum (crossing) time $\tau_\text{min}$, normalised by the undisturbed crossing time $h/V_1$, shows similar trends.
At large $Fr_1$, the crossing time is practically unity, which is consistent with the observation that $V_\text{min} / V_1$ is close to unity.
When $Fr_1$ is close to $Fr_{1;\text{lev}}$, the residence time in the layer increases dramatically.
Again, this is not surprising as this is where $V_\text{min}/V_1 \ll 1$, implying significant decelerations in the interface layer and thus large residence times.
This effect can be observed to become stronger as $\rho_p/\rho_1$ becomes larger.

\section{Concluding remarks}\label{sec:summary}

In this paper we extended the understanding of the problem first formulated and studied in~\citet[][SMF]{Fernando1999} -- the observed increase in particle settling times due to the additional force created by particles in variable density environments. 
We demonstrated, through a set of experiments in which the particle Reynolds and Froude numbers were varied not only through size, but also through particle density, that the current model for the stratification force in the form of a drag-type relation is incomplete.
It was shown that for the current data as well as a subset of the SMF data, the existence of a minimum velocity is associated with the particle exiting the density interface.
This is at variance with the hypothesis of wake rupturing  being responsible for the minimum, and further work is needed to establish the specific physical mechanisms that are responsible for this effect.

It was shown that none of the parameterisations reported in the literature that capture $F_S$ in linearly stratified environments were able to capture the transient behaviour of $F_S$. Therefore, a simple theoretical model was derived which captures adequately the behaviour of all particle types.
The conceptual picture of the  model is based on the idea of an effective volume of a fluid of different density which remains attached to the particle, and adds its buoyancy to the resisting forces. Due to this additional buoyancy-related force the particle slows down as it crosses the interface and also as leaves the interface layer and settles (or rises) in the second homogeneous density layer. 
The model is shown to predict the trajectories of not only all our particles, but also of digitised data from SMF, which is a demonstration of its predictive capabilities. 

The model shows that that this problem depends on four dimensionless quantities, namely the entrance Froude number $Fr_1$, the entrance Reynolds number $Re_1$, the density ratio $\rho_p/\rho_1$, and the relative density interface thickness $h/a$. 
For fluids in which the viscosity is different between the layers, as was the case for our index-matched experiments, the problem additionally depends on the viscosity ratio $\nu_2/\nu_1$.
Even though the model was not developed with particle levitation in mind, it does provide predictions for when this phenomenon might occur.

The results presented in this paper were obtained with  relatively sharp density interfaces.
For much thicker interfaces, one would expect different behaviour; for example our observation that the minimum velocity coincides with exiting the interface will require revision for much thicker interfaces, as the minimal velocity might not occur or occur elsewhere \citep{Doostmohammadi2014, hanazaki2009}.
Appropriate modelling of 
fluid dynamics around the sphere needs to be implemented to understand the processes in the caudal wake, incorporating buoyancy effects resulting from the density difference between the fluid around the particle and the surrounding fluid.

Furthermore, the physical mechanism that causes the behaviour observed in this study should be studied in detail. 
Indeed, by matching the refractive indices of the two fluids, it was impossible to carry out shadowgraphy and other detailed flow visualisations that would have helped in guiding the modelling. 
Accurate PLIF and density measurements are required with a larger zoom to improve the resolution of the interface detection and reveal physical phenomena during particle entrance.
It would be interesting to apply the decomposition technique introduced in \cite{Zhang2019} to the cases under consideration here.
This would allow a systematic investigation into what causes the stratification force, and whether it is also the effects of vorticity which are dominant as they are in the linearly stratified situation.

\section*{Acknowledgements}
This work received funding from the European Union
Horizon 2020 Research and Innovation Programme under the Marie Sklodowska-Curie Actions, Grant Agreement
No. 675675 and Israel Science Foundation, Grant Agreement 945/15.

\appendix

\clearpage

\section{Properties of all the particles}

In the appendix we report all the reconstructed values of density, diameter with the characteristic Froude, Reynolds and Archimedes numbers associated to each sphere released in the experiments. 
The Archimedes number ($A_{r_{i}}$) is calculated according to definition:
\begin{equation}
A_{r_{i}} = g a^3 \rho_{i} (\rho_p - \rho_{i})/\mu_i^2
\end{equation}
where $i$ assumes $i=1,2$ respectively for the top and bottom layer.

\begin{table}
\centering
\caption{Reconstructed properties of the particles of type P1.}
\label{tab:p1_table}
\begin{tabular}{ccccccccc}
\hline
\textbf{Id} & \textbf{$\rho_p$ [kg/m$^{3}$]} & \textbf{$a$ [$\mu$ m]}   & \textbf{$Re_1$}& \textbf{$Re_2$}& \textbf{$Fr_1$} & \textbf{$Fr_2$}& \textbf{$Ar_1$}& \textbf{$Ar_2$}\\ \hline

\textbf{1}            &   1083     &   839    &   10.1   &  10.3    &  3.2  &   2.3    &  311   &  321   \\ \hline
\textbf{2}            &   1042     &   920    &   8.6    &  4.8     &  2.3  &   0.9    &  253   &  126   \\ \hline
\textbf{3}            &   1041     &   945    &   9.0    &  5.0     &  2.2  &   0.9    &  271   &  129  \\ \hline
\textbf{4}            &   1059     &   878    &   9.2    &  7.6     &  2.7  &   1.5    &  278   &  219   \\ \hline
\textbf{5}            &   1035     &   914    &   7.7    &  3.0     &  2.1  &   0.6    &  222   &  72  \\ \hline
\textbf{6}            &   1040     &   852    &   6.7    &  3.7     &  2.1  &   0.8    &  196   &  91   \\ \hline
\textbf{7}            &   1035     &   859    &   7.6    &  2.7     &  2.0  &   0.5    &  186   &  63   \\ \hline
\textbf{8}            &   1033     &   919    &   7.9    &  2.6     &  2.0  &   0.5    &  219   &  61  \\ \hline
\textbf{9}            &   1036     &   918    &   10.3   &  3.4     &  2.1  &   0.6    &  229   &  82   \\ \hline
\textbf{10}           &   1061     &   915    &   13.6   &  8.7     &  2.7  &   1.8    &  320   &  259  \\ \hline
\textbf{11}           &   1105     &   896    &   9.7    &  15.4    &  3.8  &   3.0    &  457   &  541   \\ \hline
\textbf{12}           &   1048     &   942    &   8.3    &  6.6     &  2.5  &   1.1    &  298   &  184    \\ \hline
\textbf{13}           &   1040     &   914    &   9.3    &  4.4     &  2.2  &   0.8    &  243   &  112  \\ \hline
\textbf{14}           &   1052     &   907    &   9.3    &  6.9     &  2.5  &   1.2    &  282   &  194   \\ \hline
\textbf{15}           &   1055     &   896    &   7.3    &  7.1     &  2.6  &   1.4    &  279   &  202   \\ \hline
\textbf{16}           &   1037     &   893    &   7.5    &  3.4     &  2.1  &   0.7    &  215   &  82   \\ \hline
\textbf{17}           &   1045     &   908    &   8.6    &  5.3     &  2.3  &   1.0    &  254   &  141   \\ \hline
\textbf{18}           &   1077     &   891    &   11.1   &  11.0    &  3.1  &   2.2    &  354   &  349   \\ \hline
\end{tabular}
\end{table}

\begin{table}
\centering
\caption{Reconstructed properties of the particles of type P2.}
\label{tab:p2_table}
\begin{tabular}{ccccccccc}
\hline
\textbf{Id} & \textbf{$\rho_p$ [kg/m$^{3}$]} & \textbf{$a$ [$\mu$ m]}   & \textbf{$Re_1$}& \textbf{$Re_2$}& \textbf{$Fr_1$} & \textbf{$Fr_2$}& \textbf{$Ar_1$}& \textbf{$Ar_2$}\\ \hline

\textbf{1}            &   1194     &   478    &   4.6    &  6.3     &  4.5  &   4.6    &  117   &  173   \\ \hline
\textbf{2}            &   1208     &   438    &   3.9    &  5.4     &  4.5  &   4.5    &  96    &  144   \\ \hline
\textbf{3}            &   1235     &   445    &   4.4    &  6.3     &  5.0  &   5.1    &  112   &  173  \\ \hline
\textbf{4}            &   1198     &   453    &   4.0    &  5.6     &  4.5  &   4.4    &  101   &  150   \\ \hline
\textbf{5}            &   1207     &   434    &   3.8    &  5.3     &  4.5  &   4.5    &  93    &  139  \\ \hline
\textbf{6}            &   1257     &   440    &   4.6    &  6.6     &  5.3  &   5.5    &  117   &  184   \\ \hline
\textbf{7}            &   1207     &   454    &   4.2    &  5.9     &  4.6  &   4.6    &  106   &  159   \\ \hline
\textbf{8}            &   1238     &   424    &   4.4    &  6.3     &  5.0  &   5.1    &  111   &  172  \\ \hline
\textbf{9}            &   1170     &   480    &   4.2    &  5.6     &  4.1  &   4.1    &  105   &  149   \\ \hline
\textbf{10}           &   1238     &   458    &   4.8    &  6.8     &  5.1  &   5.2    &  123   &  191  \\ \hline
\textbf{11}           &   1200     &   463    &   4.3    &  6.0     &  4.5  &   4.7    &  109   &  162   \\ \hline
\end{tabular}
\end{table}

\begin{table}
\centering
\caption{Reconstructed properties of the particles of type P3.}
\label{tab:p3_table}
\begin{tabular}{ccccccccc}
\hline
\textbf{Id} & \textbf{$\rho_p$ [kg/m$^{3}$]} & \textbf{$a$ [$\mu$ m]}   & \textbf{$Re_1$}& \textbf{$Re_2$}& \textbf{$Fr_1$} & \textbf{$Fr_2$}& \textbf{$Ar_1$}& \textbf{$Ar_2$}\\ \hline

\textbf{1}            &   2558     &   733    &   53    &  81     &  24  &   26    &  3062   &  5643   \\ \hline
\textbf{2}            &   2503     &   787    &   60    &  90     &  23  &   25    &  3654   & 6724    \\ \hline
\textbf{3}            &   2395     &   774    &   55    &  83     &  22  &   24    &  3236   &  5940  \\ \hline
\textbf{4}            &   2361     &   792    &   57    &  86     &  22  &   23    &  3383   &  6206   \\ \hline
\textbf{5}            &   2278     &   902    &   71    &  106    &  21  &   22    &  4697   & 8594   \\ \hline
\textbf{6}            &   2474     &   838    &   67    &  101    &  23  &   25    &  4333   &  7969   \\ \hline
\textbf{7}            &   2347     &   871    &   69    &  103    &  22  &   23    &  4454   &  8167   \\ \hline
\textbf{8}            &   2619     &   698    &   49    &  75     &  25  &   26    &  2746   &  5066  \\ \hline
\textbf{9}            &   2304     &   876    &   68    &  102    &  21  &   23    &  4389   &  8038   \\ \hline
\textbf{10}           &   2779     &   688    &   51    &  78     &  26  &   28    &  2886   &  5338  \\ \hline
\textbf{11}           &   2619     &   769    &   60    &  91     &  25  &   26    &  3674   &  6777   \\ \hline
\end{tabular}
\end{table}

\begin{table}
\centering
\caption{Reconstructed properties of the particles of type P4.}
\label{tab:p4_table}
\begin{tabular}{ccccccccc}
\hline
\textbf{Id} & \textbf{$\rho_p$ [kg/m$^{3}$]} & \textbf{$a$ [$\mu$ m]}   & \textbf{$Re_1$}& \textbf{$Re_2$}& \textbf{$Fr_1$} & \textbf{$Fr_2$}& \textbf{$Ar_1$}& \textbf{$Ar_2$}\\ \hline

\textbf{1}            &   960     &   877    &   8.2    &  4.2    &  2.3  &   0.8    &  70   &  375   \\ \hline
\textbf{2}            &   959     &   821    &   7.2    &  3.7    &  2.3  &   0.8    &  61   &  315   \\ \hline
\textbf{3}            &   956     &   777    &   6.5    &  3.5    &  2.3  &   0.9    &  57   &  277  \\ \hline
\textbf{4}            &   959     &   820    &   7.1    &  3.6    &  2.3  &   0.8    &  59   &  311   \\ \hline
\textbf{5}            &   968     &   894    &   7.7    &  2.9    &  2.1  &   0.5    &  45   &  343  \\ \hline
\textbf{6}            &   956     &   823    &   7.5    &  4.2    &  2.4  &   0.9    &  69   &  333   \\ \hline
\textbf{7}            &   962     &   856    &   7.6    &  3.6    &  2.2  &   0.7    &  58   &  336   \\ \hline
\textbf{8}            &   966     &   872    &   7.5    &  3.1    &  2.1  &   0.6    &  50   &  333  \\ \hline
\textbf{9}            &   965     &   851    &   7.2    &  3.1    &  2.0  &   0.6    &  50   &  317   \\ \hline
\textbf{10}           &   970     &   884    &   7.3    &  2.6    &  2.1  &   0.5    &  40   &  324  \\ \hline
\textbf{11}           &   964     &   819    &   6.7    &  3.0    &  2.2  &   0.6    &  47   &  287   \\ \hline
\textbf{12}           &   962     &   852    &   7.5    &  3.6    &  2.3  &   0.7    &  59   &  335    \\ \hline
\textbf{13}           &   961     &   800    &   6.6    &  3.2    &  2.2  &   0.7    &  52   &  282  \\ \hline
\textbf{14}           &   961     &   873    &   8.0    &  4.0    &  2.3  &   0.8    &  62   &  365   \\ \hline
\textbf{15}           &   966     &   862    &   7.3    &  3.0    &  2.1  &   0.6    &  49   &  323   \\ \hline

\end{tabular}
\end{table}

\begin{table}
\centering
\caption{Reconstructed properties of the particles of \cite{Fernando1999}.}
\label{tab:Fernando_table}
\begin{tabular}{ccccccccc}
\hline
\textbf{Id} & \textbf{$\rho_p$ [kg/m$^{3}$]} & \textbf{$a$ [$\mu$ m]}   & \textbf{$Re_1$}& \textbf{$Re_2$}& \textbf{$Fr_1$} & \textbf{$Fr_2$}& \textbf{$Ar_1$}& \textbf{$Ar_2$}\\ \hline
\textbf{1} & 2500  & 380	&  	 23& 	21.7& 	26  & 	24.4& 	774& 	732 \\ \hline
\textbf{2} & 1052  & 1000	&  10.1& 	2.1 & 	1.6 & 	0.3 & 	305& 	64 \\ \hline
\textbf{3} & 1069  & 718	& 	6.3& 	1.2 & 	2   &	0.3 & 	172& 	34 \\ \hline
\textbf{4} & 1068  & 737	& 	6.9& 	1.2 & 	2   &	0.4 & 	183& 	33 \\ \hline
\textbf{5} & 1052  & 856	& 	6.9& 	1.4 & 	1.5 & 	0.3 & 	191& 	40 \\ \hline
\textbf{6} & 1050  & 747	& 	4.7& 	1.3 & 	1.3 & 	0.4 & 	119& 	34 \\ \hline
\textbf{7} & 1100  & 483	& 	3.5& 	1.0 & 	2.5 & 	1.2 & 	85 & 	42 \\ \hline
\textbf{8} & 1083  & 522	& 	3.5& 	1.2 & 	2.1 & 	0.7 & 	85 & 	31 \\ \hline
\textbf{9} & 1051  & 678	& 	3.8& 	0.9 & 	1.3 & 	0.3 & 	92 & 	23 \\ \hline
\textbf{10} & 1052 & 622	& 	3.1& 	0.6 & 	1.3 & 	0.2 & 	73 & 	15 \\ \hline
\textbf{11}&  1049 & 644	& 	3.1& 	1.0 & 	1.2 & 	0.4 & 	73 & 	24 \\ \hline
\textbf{12} & 1051 & 522    & 	1.9& 	0.4 & 	1.1 & 	0.2 & 	42 & 	10 \\ \hline
\textbf{13} & 1050 & 373	& 	0.7& 	0.2 & 	0.8 & 	0.2 & 	14 & 	4 \\ \hline
\end{tabular}
\end{table}

\bibliographystyle{model2-names}
\bibliography{Particles_crossing}

\begin{thebibliography}{31}
\expandafter\ifx\csname natexlab\endcsname\relax\def\natexlab#1{#1}\fi
\providecommand{\url}[1]{\texttt{#1}}
\providecommand{\href}[2]{#2}
\providecommand{\path}[1]{#1}
\providecommand{\DOIprefix}{doi:}
\providecommand{\ArXivprefix}{arXiv:}
\providecommand{\URLprefix}{URL: }
\providecommand{\Pubmedprefix}{pmid:}
\providecommand{\doi}[1]{\href{http://dx.doi.org/#1}{\path{#1}}}
\providecommand{\Pubmed}[1]{\href{pmid:#1}{\path{#1}}}
\providecommand{\bibinfo}[2]{#2}
\ifx\xfnm\relax \def\xfnm[#1]{\unskip,\space#1}\fi
\bibitem[{Abaid and Adalsteinsson(2004)}]{Adalsteinsson2004}
\bibinfo{author}{Abaid, N.}, \bibinfo{author}{Adalsteinsson, D.},
  \bibinfo{year}{2004}.
\newblock \bibinfo{title}{{An internal splash: Levitation of falling spheres in
  stratified fluids}}.
\newblock \bibinfo{journal}{Physics of Fluids} \bibinfo{volume}{16},
  \bibinfo{pages}{1567}.
\bibitem[{Alahyari and Longmire(1994)}]{alahyari1994}
\bibinfo{author}{Alahyari, A.}, \bibinfo{author}{Longmire, E.},
  \bibinfo{year}{1994}.
\newblock \bibinfo{title}{Particle image velocimetry in a variable density
  flow: application to a dynamically evolving microburst}.
\newblock \bibinfo{journal}{Experiments in Fluids} \bibinfo{volume}{17},
  \bibinfo{pages}{434--440}.
\bibitem[{Ardekani and Stocker(2010)}]{ardekani2010}
\bibinfo{author}{Ardekani, A.}, \bibinfo{author}{Stocker, R.},
  \bibinfo{year}{2010}.
\newblock \bibinfo{title}{Stratlets: Low {R}eynolds number point-force
  solutions in a stratified fluid}.
\newblock \bibinfo{journal}{Physical Review Letters} \bibinfo{volume}{105},
  \bibinfo{pages}{084502}.
\bibitem[{Burd and Jackson(2009)}]{burd2009}
\bibinfo{author}{Burd, A.}, \bibinfo{author}{Jackson, G.},
  \bibinfo{year}{2009}.
\newblock \bibinfo{title}{Particle aggregation}.
\newblock \bibinfo{journal}{Annual Review of Marine Science}
  \bibinfo{volume}{1}, \bibinfo{pages}{65--90}.
\bibitem[{Camassa et~al.(2010)Camassa, Falcon, Lin, McLaughlin and
  Mykins}]{Camassa2010}
\bibinfo{author}{Camassa, R.}, \bibinfo{author}{Falcon, C.},
  \bibinfo{author}{Lin, J.}, \bibinfo{author}{McLaughlin, R.},
  \bibinfo{author}{Mykins, N.}, \bibinfo{year}{2010}.
\newblock \bibinfo{title}{A first-principle predictive theory for a sphere
  falling through sharply stratified fluid at low {R}eynolds number}.
\newblock \bibinfo{journal}{Journal of Fluid Mechanics} \bibinfo{volume}{664},
  \bibinfo{pages}{436--465}.
\bibitem[{Camassa et~al.(2009)Camassa, Falcon, Lin, McLaughlin and
  Parker}]{Camassa2009}
\bibinfo{author}{Camassa, R.}, \bibinfo{author}{Falcon, C.},
  \bibinfo{author}{Lin, J.}, \bibinfo{author}{McLaughlin, R.},
  \bibinfo{author}{Parker, R.}, \bibinfo{year}{2009}.
\newblock \bibinfo{title}{Prolonged residence times for particles settling
  through stratified miscible fluids in the {S}tokes regime}.
\newblock \bibinfo{journal}{Physics of Fluids} \bibinfo{volume}{21},
  \bibinfo{pages}{031702}.
\bibitem[{Camassa et~al.(2013)Camassa, Khatri, McLaughlin, Prairie, White and
  Yu}]{Camassa2013}
\bibinfo{author}{Camassa, R.}, \bibinfo{author}{Khatri, S.},
  \bibinfo{author}{McLaughlin, R.}, \bibinfo{author}{Prairie, J.},
  \bibinfo{author}{White, B.}, \bibinfo{author}{Yu, S.}, \bibinfo{year}{2013}.
\newblock \bibinfo{title}{Retention and entrainment effects: Experiments and
  theory for porous spheres settling in sharply stratified fluids}.
\newblock \bibinfo{journal}{Physics of Fluids} \bibinfo{volume}{25},
  \bibinfo{pages}{081701}.
\bibitem[{Candelier et~al.(2014)Candelier, Mehaddi and
  Vauquelin}]{Candelier2014}
\bibinfo{author}{Candelier, F.}, \bibinfo{author}{Mehaddi, R.},
  \bibinfo{author}{Vauquelin, O.}, \bibinfo{year}{2014}.
\newblock \bibinfo{title}{The history force on a small particle in a linearly
  stratified fluid}.
\newblock \bibinfo{journal}{Journal of Fluid Mechanics} \bibinfo{volume}{749},
  \bibinfo{pages}{184--200}.
\bibitem[{Clift et~al.(2005)Clift, Grace and Weber}]{Clift2005}
\bibinfo{author}{Clift, R.}, \bibinfo{author}{Grace, J.R.},
  \bibinfo{author}{Weber, M.}, \bibinfo{year}{2005}.
\newblock \bibinfo{title}{Bubbles, drops, and particles}.
\newblock \bibinfo{publisher}{Courier Corporation}.
\bibitem[{Doostmohammadi et~al.(2014)Doostmohammadi, Dabiri and
  Ardekani}]{Doostmohammadi2014}
\bibinfo{author}{Doostmohammadi, A.}, \bibinfo{author}{Dabiri, S.},
  \bibinfo{author}{Ardekani, A.}, \bibinfo{year}{2014}.
\newblock \bibinfo{title}{A numerical study of the dynamics of a particle
  settling at moderate reynolds numbers in a linearly stratified fluid}.
\newblock \bibinfo{journal}{Journal of Fluid Mechanics} \bibinfo{volume}{750},
  \bibinfo{pages}{5--32}.
\bibitem[{Eames and Hunt(1997)}]{eames1997}
\bibinfo{author}{Eames, I.}, \bibinfo{author}{Hunt, J.}, \bibinfo{year}{1997}.
\newblock \bibinfo{title}{Inviscid flow around bodies moving in weak density
  gradients without buoyancy effects}.
\newblock \bibinfo{journal}{Journal of Fluid Mechanics} \bibinfo{volume}{353},
  \bibinfo{pages}{331--355}.
\bibitem[{Geller et~al.(1986)Geller, Lee and Leal}]{geller1986}
\bibinfo{author}{Geller, A.}, \bibinfo{author}{Lee, S.}, \bibinfo{author}{Leal,
  L.}, \bibinfo{year}{1986}.
\newblock \bibinfo{title}{The creeping motion of a spherical particle normal to
  a deformable interface}.
\newblock \bibinfo{journal}{Journal of Fluid Mechanics} \bibinfo{volume}{169},
  \bibinfo{pages}{27--69}.
\bibitem[{Hanazaki et~al.(2009)Hanazaki, Konishi and Okamura}]{hanazaki2009}
\bibinfo{author}{Hanazaki, H.}, \bibinfo{author}{Konishi, K.},
  \bibinfo{author}{Okamura, T.}, \bibinfo{year}{2009}.
\newblock \bibinfo{title}{Schmidt-number effects on the flow past a sphere
  moving vertically in a stratified diffusive fluid}.
\newblock \bibinfo{journal}{Physics of Fluids} \bibinfo{volume}{21},
  \bibinfo{pages}{026602}.
\bibitem[{Kok(2011)}]{kok2011}
\bibinfo{author}{Kok, J.F.}, \bibinfo{year}{2011}.
\newblock \bibinfo{title}{A scaling theory for the size distribution of emitted
  dust aerosols suggests climate models underestimate the size of the global
  dust cycle}.
\newblock \bibinfo{journal}{Proceedings of the National Academy of Sciences}
  \bibinfo{volume}{108}, \bibinfo{pages}{1016--1021}.
\bibitem[{MacIntyre et~al.(1995)MacIntyre, Alldredge and
  Gotschalk}]{macintyre1995}
\bibinfo{author}{MacIntyre, S.}, \bibinfo{author}{Alldredge, A.L.},
  \bibinfo{author}{Gotschalk, C.C.}, \bibinfo{year}{1995}.
\newblock \bibinfo{title}{Accumulation of marine snow at density
  discontinuities in the water column}.
\newblock \bibinfo{journal}{Limnology and Oceanography} \bibinfo{volume}{40},
  \bibinfo{pages}{449--468}.
\bibitem[{Magnaudet(2020)}]{Magnaudet2020}
\bibinfo{author}{Magnaudet, J.and~Mercier, M.}, \bibinfo{year}{2020}.
\newblock \bibinfo{title}{Particles, drops and bubbles moving across sharp
  interfaces and stratified layers}.
\newblock \bibinfo{journal}{Ann .Rev. Fluid Mech.} \bibinfo{volume}{in press}.
\bibitem[{Maxey and Riley(1983)}]{maxey1983}
\bibinfo{author}{Maxey, M.}, \bibinfo{author}{Riley, J.}, \bibinfo{year}{1983}.
\newblock \bibinfo{title}{Equation of motion for a small rigid sphere in a
  nonuniform flow}.
\newblock \bibinfo{journal}{The Physics of Fluids} \bibinfo{volume}{26},
  \bibinfo{pages}{883--889}.
\bibitem[{Odar and Hamilton(1964)}]{odar1964}
\bibinfo{author}{Odar, F.}, \bibinfo{author}{Hamilton, W.},
  \bibinfo{year}{1964}.
\newblock \bibinfo{title}{Forces on a sphere accelerating in a viscous fluid}.
\newblock \bibinfo{journal}{Journal of Fluid Mechanics} \bibinfo{volume}{18},
  \bibinfo{pages}{302--314}.
\bibitem[{Okino et~al.(2017)Okino, Akiyama and Hanazaki}]{okino2017}
\bibinfo{author}{Okino, S.}, \bibinfo{author}{Akiyama, S.},
  \bibinfo{author}{Hanazaki, H.}, \bibinfo{year}{2017}.
\newblock \bibinfo{title}{Velocity distribution around a sphere descending in a
  linearly stratified fluid}.
\newblock \bibinfo{journal}{Journal of Fluid Mechanics} \bibinfo{volume}{826},
  \bibinfo{pages}{759--780}.
\bibitem[{Pierson and Magnaudet(2018a)}]{magnaudet2018}
\bibinfo{author}{Pierson, J.}, \bibinfo{author}{Magnaudet, J.},
  \bibinfo{year}{2018}a.
\newblock \bibinfo{title}{Inertial settling of a sphere through an interface.
  {P}art 1. {F}rom sphere flotation to wake fragmentation}.
\newblock \bibinfo{journal}{Journal of Fluid Mechanics} \bibinfo{volume}{835},
  \bibinfo{pages}{762--807}.
\bibitem[{Pierson and Magnaudet(2018b)}]{magnaudet2018b}
\bibinfo{author}{Pierson, J.}, \bibinfo{author}{Magnaudet, J.},
  \bibinfo{year}{2018}b.
\newblock \bibinfo{title}{Inertial settling of a sphere through an interface.
  {P}art 2. {S}phere and tail dynamics}.
\newblock \bibinfo{journal}{Journal of Fluid Mechanics} \bibinfo{volume}{835},
  \bibinfo{pages}{808--851}.
\bibitem[{Prairie et~al.(2013)Prairie, Ziervogel, Arnosti, Camassa, Falcon,
  Khatri, McLaughlin, White and Yu}]{prairie2013}
\bibinfo{author}{Prairie, J.}, \bibinfo{author}{Ziervogel, K.},
  \bibinfo{author}{Arnosti, C.}, \bibinfo{author}{Camassa, R.},
  \bibinfo{author}{Falcon, C.}, \bibinfo{author}{Khatri, S.},
  \bibinfo{author}{McLaughlin, R.}, \bibinfo{author}{White, B.},
  \bibinfo{author}{Yu, S.}, \bibinfo{year}{2013}.
\newblock \bibinfo{title}{Delayed settling of marine snow at sharp density
  transitions driven by fluid entrainment and diffusion-limited retention}.
\newblock \bibinfo{journal}{Marine Ecology Progress Series}
  \bibinfo{volume}{487}, \bibinfo{pages}{185--200}.
\bibitem[{Smith et~al.(1992)Smith, Simon, Alldredge and Azam}]{smith1992}
\bibinfo{author}{Smith, D.}, \bibinfo{author}{Simon, M.},
  \bibinfo{author}{Alldredge, A.}, \bibinfo{author}{Azam, F.},
  \bibinfo{year}{1992}.
\newblock \bibinfo{title}{Intense hydrolytic enzyme activity on marine
  aggregates and implications for rapid particle dissolution}.
\newblock \bibinfo{journal}{Nature} \bibinfo{volume}{359},
  \bibinfo{pages}{139}.
\bibitem[{Srdi{\'c}-Mitrovi{\'c} et~al.(1999)Srdi{\'c}-Mitrovi{\'c}, Mohamed
  and Fernando}]{Fernando1999}
\bibinfo{author}{Srdi{\'c}-Mitrovi{\'c}, A.}, \bibinfo{author}{Mohamed, N.},
  \bibinfo{author}{Fernando, H.}, \bibinfo{year}{1999}.
\newblock \bibinfo{title}{Gravitational settling of particles through density
  interfaces}.
\newblock \bibinfo{journal}{Journal of Fluid Mechanics} \bibinfo{volume}{381},
  \bibinfo{pages}{175--198}.
\bibitem[{Torres et~al.(2000)Torres, Hanazaki, Ochoa, Castillo and
  Van~Woert}]{torres2000}
\bibinfo{author}{Torres, C.}, \bibinfo{author}{Hanazaki, H.},
  \bibinfo{author}{Ochoa, J.}, \bibinfo{author}{Castillo, J.},
  \bibinfo{author}{Van~Woert, M.}, \bibinfo{year}{2000}.
\newblock \bibinfo{title}{Flow past a sphere moving vertically in a stratified
  diffusive fluid}.
\newblock \bibinfo{journal}{Journal of Fluid Mechanics} \bibinfo{volume}{417},
  \bibinfo{pages}{211--236}.
\bibitem[{Turco et~al.(1983)Turco, Toon, Ackerman, Pollack and
  Sagan}]{turco1983}
\bibinfo{author}{Turco, R.P.}, \bibinfo{author}{Toon, O.},
  \bibinfo{author}{Ackerman, T.}, \bibinfo{author}{Pollack, J.B.},
  \bibinfo{author}{Sagan, C.}, \bibinfo{year}{1983}.
\newblock \bibinfo{title}{Nuclear winter: Global consequences of multiple
  nuclear explosions}.
\newblock \bibinfo{journal}{Science} \bibinfo{volume}{222},
  \bibinfo{pages}{1283--1292}.
\bibitem[{Verso et~al.(2017)Verso, van Reeuwijk and Liberzon}]{verso2017}
\bibinfo{author}{Verso, L.}, \bibinfo{author}{van Reeuwijk, M.},
  \bibinfo{author}{Liberzon, A.}, \bibinfo{year}{2017}.
\newblock \bibinfo{title}{Steady state model and experiment for an oscillating
  grid turbulent two-layer stratified flow}.
\newblock \bibinfo{journal}{Physical Review Fluids} \bibinfo{volume}{2},
  \bibinfo{pages}{104605}.
\bibitem[{White(1974)}]{white1974}
\bibinfo{author}{White, F.}, \bibinfo{year}{1974}.
\newblock \bibinfo{title}{Viscous fluid flow}.
\newblock \bibinfo{publisher}{New York, McGraw-Hill}.
\bibitem[{Yick et~al.(2009)Yick, Torres, Peacock and Stocker}]{yick2009}
\bibinfo{author}{Yick, K.}, \bibinfo{author}{Torres, C.},
  \bibinfo{author}{Peacock, T.}, \bibinfo{author}{Stocker, R.},
  \bibinfo{year}{2009}.
\newblock \bibinfo{title}{Enhanced drag of a sphere settling in a stratified
  fluid at small reynolds numbers}.
\newblock \bibinfo{journal}{Journal of Fluid Mechanics} \bibinfo{volume}{632},
  \bibinfo{pages}{49--68}.
\bibitem[{Zhang et~al.(2019)Zhang, Mercier and Magnaudet}]{Zhang2019}
\bibinfo{author}{Zhang, J.}, \bibinfo{author}{Mercier, M.},
  \bibinfo{author}{Magnaudet, J.}, \bibinfo{year}{2019}.
\newblock \bibinfo{title}{Core mechanisms of drag enhancement on bodiies
  settling in a stratified fluid}.
\newblock \bibinfo{journal}{Journal of Fluid Mechanics} \bibinfo{volume}{875},
  \bibinfo{pages}{622--656}.
\bibitem[{Zvirin and Chadwick(1975)}]{zvirin1975}
\bibinfo{author}{Zvirin, Y.}, \bibinfo{author}{Chadwick, R.},
  \bibinfo{year}{1975}.
\newblock \bibinfo{title}{Settling of an axially symmetric body in a viscous
  stratified fluid}.
\newblock \bibinfo{journal}{International Journal of Multiphase Flow}
  \bibinfo{volume}{1}, \bibinfo{pages}{743--752}.

\end{thebibliography}

\end{document}